\documentclass{IEEEtran}
\usepackage{cite}
\usepackage{amsmath,amssymb,amsfonts}
\usepackage{algorithmic}
\usepackage{graphicx}
\usepackage{textcomp}
\usepackage{tikz}
\usepackage{graphicx} 
\usepackage{layout}
\usepackage{array}
\usepackage{amsmath}
\usepackage{amssymb}
\usepackage{color}
\usepackage{epstopdf}
\usepackage{algorithm} 
\usepackage{algorithmic}
\usepackage{import}
\usepackage{tikz}
\usepackage{pgfplots}
\usepackage{tikz,tikz-3dplot}
\usepackage{bm}
\usepackage{subfig}

\usepackage{array}
\newcolumntype{P}[1]{>{\centering\arraybackslash}p{#1}}
\newcolumntype{M}[1]{>{\centering\arraybackslash}m{#1}}
\usepackage{booktabs}

\usepackage{tabularx, booktabs, makecell, caption}
    \usepackage{siunitx}

\usepackage{accents}

\usepackage{amsmath}
\DeclareMathOperator\asinh{asinh}

\makeatletter
\newcommand{\thickhline}{%
    \noalign {\ifnum 0=`}\fi \hrule height 1.5pt
    \futurelet \reserved@a \@xhline
}
\newcolumntype{"}{@{\hskip\tabcolsep\vrule width 1pt\hskip\tabcolsep}}
\makeatother

\DeclareMathOperator\erfc{erfc}

\usetikzlibrary{arrows.meta}
\usetikzlibrary{calc}
\tikzset{axis line style/.style={thin, gray, -stealth}}

\def\BibTeX{{\rm B\kern-.05em{\sc i\kern-.025em b}\kern-.08em
    T\kern-.1667em\lower.7ex\hbox{E}\kern-.125emX}}
\begin{document}

\title{A Broadband Multipole Method for \\ Accelerated Mutual Coupling Analysis of \\ Large Irregular Arrays Including Rotated Antennas}

\author{Quentin~Gueuning, Eloy de Lera Acedo, Anthony Keith Brown, \emph{Life Senior Member, IEEE},
        Christophe~Craeye, \emph{Senior Member, IEEE} and Oscar O'Hara,
\thanks{Q. Gueuning, E. de Lera Acedo and Oscar O'Hara are with the Astrophysics Group, Cavendish
Laboratory, University of Cambridge, UK and also with the Kavli Institute for Cosmology, Cavendish
Laboratory, University of Cambridge, UK.}
\thanks{A. K. Brown is with the School of Electronic Engineering and
Computer Science, Queen Mary University of London, London, UK
and also the University of Manchester, UK.}
\thanks{C. Craeye is with the Institute of Information and Communication Technologies, Electronics and
Applied Mathematics (ICTEAM), Universit\'e Catholique de Louvain, Louvain-la-Neuve, Belgium.}
}

\markboth{IEEE Trans. Antennas Propag., vol. XXX, no. XXX }%
{Shell \MakeLowercase{\textit{et al.}}: Bare Demo of IEEEtran.cls for IEEE Journals}
\maketitle
\begin{abstract}
We present a numerical method for the analysis of mutual coupling effects in large, dense and irregular arrays with identical antennas. Building on the Method of Moments (MoM), our technique employs a Macro Basis Function (MBF) approach for rapid direct inversion of the MoM impedance matrix. To expedite the reduced matrix filling, we propose an extension of the Steepest-Descent Multipole expansion which remains numerically stable and efficient across a wide bandwidth. This broadband multipole-based approach is well suited to quasi-planar problems and requires only the pre-computation of each MBF's complex patterns, resulting in low antenna-dependent pre-processing costs. The method also supports arrays with arbitrarily rotated antennas at low additional cost. A simulation of all embedded element patterns of irregular arrays of 256 complex log-periodic antennas completes in just 10 minutes per frequency point on a current laptop, with an additional minute per new layout.
\end{abstract}

\begin{IEEEkeywords}
Steepest Descent Path, Multipole Method, Macro Basis Functions, Mutual Coupling, Broadband, Wideband, Quasi-Planar, Antenna Arrays, Square Kilometer Array. \textcolor{blue}{This work has been submitted to the IEEE for possible publication. Copyright may be transferred without notice, after which this version may no longer be accessible.}
\end{IEEEkeywords}

\section{Introduction}
Mutual coupling (MC) refers to the electromagnetic interaction between antennas in an array, causing each element to behave differently, depending on its environment. The MC effect manifests itself as angular and frequency variations in the radiated far fields, known as embedded element patterns (EEPs), among elements in the array and in the array impedance matrix. When not carefully accounted for by design, MC can cause blind spots in radiation patterns \cite{Virone} or impedance mismatches \cite{Bird} between antennas and first-stage amplifiers. 
For instance, positioning antennas irregularly \cite{Razavi-Ghods} or sequentially rotating them \cite{Huang,Smolders,Carozzi} can help randomize MC effects. This mutual interaction can be significant in wideband arrays due to the inherent antennas support for higher-order current modes. Given the widespread use of large wideband phased array systems \cite{Latha} in radar, space communication, biomedical imaging, radio astronomy, and mobile communications, accurate modeling of each element's response in the presence of MC is important and enables the mitigation of undesired effects through the optimization of antenna geometry, array layout, and the integration with front-end electronics. 

Analyzing MC requires solving Maxwell's equations using full-wave solvers capable of handling large, multi-scale problems. Wideband antennas often have intricate geometries needing hundreds \cite{Maaskant} or thousands \cite{SKALA4} of mesh elements, while array sizes can extend to hundreds of wavelengths \cite{HIRAX}. Additionally, fine frequency resolution is required to capture severely-narrowband MC effects \cite{Bolli}. General-purpose solvers based on the Method of Moments (MoM,\cite{Harrington}) often result in prohibitive computation times; that is days or weeks per frequency on large workstations, even when accelerated with the Multi-Level Fast Multipole Method (MLFMM, \cite{Rokhlin}) \cite{Davidson2}). The irregularity of the layout also prevents the use of periodic boundary conditions. In contrast, Fast Computational ElectroMagnetics (CEM) methods have been developed specifically for large irregular arrays, reducing potentially computation times to minutes/hours on a laptop \cite{Maaskant,HARP,Conradie,Ludick}.

Solving the MoM system of equations for dense and electrically large problems is costly due to filling and inverting the impedance matrix, therefore requiring acceleration techniques. For array problems with identical and disconnected antennas, direct inversion techniques \cite{Maaskant, HARP, Gonzalez, Craeyemultipole, Gurel, Matekovits, Suter, Prakash, Gueuning} are preferred over iterative ones \cite{Jandhyala, Freni, Conradie} to avoid the use of preconditioners and to allow quick solutions for each excited array port. Frameworks such as Macro-Basis Function (MBF, \cite{Suter}), Characteristic-Basis Function (CBF, \cite{Prakash}) or Synthetic Function eXpansion (SFX,\cite{Matekovits}) enable direct matrix inversion by defining current basis functions over the antenna domain, hence reducing the number of unknowns per antenna. Despite reducing inversion costs, the impedance filling time remains equivalent to that of the brute force MoM and must thus also be sped up. This can be achieved by modelling MBF interactions as a function of baselines using interpolative methods or analytical field expansions in a pre-computation step that is independent of array configuration.
One approach, devised in \cite{Craeyemultipole}, uses Rokhlin's multipole expansion \cite{Rokhlin}, but it suffers from a Low-Frequency (LF) breakdown \cite{Chew1999}, limiting its use for inter-element distances greater than about half a wavelength. Alternative schemes, such as those \cite{Maaskant}, \cite{Conradie} based on Adaptive Cross Approximation (ACA, \cite{Zhao}) and the HARmonic-Polynomial method (HARP, \cite{HARP},\cite{Gonzalez}), build MBF interaction models from a few (entries of) pre-computed elementary MoM blocks. For complex antennas meshed with many elements, these pre-computations can take hours to days hindering fast frequency sweeping or re-computation for various antenna geometries. In contrast, the multipole-based approach \cite{Craeyemultipole} only requires pre-computation of each MBFs' far-field patterns but cannot be used at short distances. This paper thus proposes an alternative efficient multipole expansion for quasi-planar structures that remains free from LF breakdown.

The Fast Multipole Method (FMM, \cite{Rokhlin}) is highly efficient in CEM, accelerating computations by factorising MoM interactions between a source and an observation group into a sum of plane-wave interactions involving the product of source and observation patterns with an analytical translation function, which depends only on the vector relative distance between groups. While effective for large relative distances, the multipole expansion loses accuracy when groups are closer than about half a wavelength due to the LF breakdown \cite{Gueuning3}. 
Intense efforts \cite{Chew, Xia, Cheng, Bogaert2008, Jian, Wallen, Wulf,Ergul} have focused on deriving a broadband multipole expansion for general 3D problems that remains accurate and efficient from low to high frequencies. One class of methods relies on a combination of LF-stable multipole \cite{Chew, Xia, Cheng} or algebraic methods such as QR decomposition \cite{Bogaert2008} at lower frequencies, then transitioning to the MLFMM at higher frequencies. Another popular approach involves spectral-domain techniques \cite{Jian, Wallen, Wulf}, which are typically formulated by evaluating the Weyl identity \cite{Chewbook} along a specific contour in the complex plane to include the evanescent part of the spectrum at low frequencies, thereby better capturing reactive near fields. However, this spectral decomposition is only numerically accurate in limited spatial sectors, thus requiring a partitioning of the observation domain with different sets of complex patterns for each sector.
To the author's best knowledge, no single broadband multipole-based expansion has been derived for general 3D problems yet. Interestingly, such expansions have been derived \cite{Chew1999, Bogaert2006,Meng} for 2D problems, effectively mitigating the LF breakdown with a simple re-normalization procedure.

\begin{figure}[t]
\centering
\resizebox {0.4\textwidth} {!} {
\begin{tikzpicture}
\draw[solid,fill=gray, fill opacity=.25] (0,0) circle (4.25);

\draw[solid,fill=white] (2.5,0) circle (1.3);

\draw[solid,fill=gray, fill opacity=.4] (2.5,0) circle (0.75);

\draw[solid,fill=gray, fill opacity=.25] (-3,0) circle (0.75);
\draw [black,->] (0,0) -- (0,1.0) node[left] {$\mathbf{\hat{y}}$};
\draw [black,->] (0,0) -- (1.0,0) node[above] {$\mathbf{\hat{x}}$};

\draw [black,->] (0.0,0,0) -- (-3,0,0) node[left] {$\bm{r} _{j}$};
\draw [black,->] (0.0,0,0) -- (2.5,0,0.0) node[right] {$\bm{r} _{i}$};


\draw [black,<->,dashed] (2.5,0) -- (2.00,-0.55) node[left] {$a$};

\draw [black,<->,dashed] (2.5,-0.75) -- (2.5,-1.3) node[below] {$P_{\min}$};

\node[text width=1.5cm] at (2.6,1.7) 
    {source group $i$};
\node[text width=1.5cm] at (-2.9,1.7) 
    {observation group $j$};
\node[text width=4cm] at (0.5,3.0) 
    {observation domain};

\draw [black,<->,dashed] (-4.25,-4.8) -- (4.25,-4.8) ;
\node[text width=2cm] at (0.5,-4.5)  {$ P_{\max}$}; 

\draw[solid,fill=gray, fill opacity=.25] (-4.25,-6.5) rectangle (4.25,-5.5);

\draw[solid,fill=gray, fill opacity=.25] (-3.75,-6.5) rectangle (-2.25,-5.5);

\draw[solid,fill=gray, fill=white] (1.2,-6.5) rectangle (3.8,-5.5);
\draw[solid,fill=gray, fill opacity=.4] (3.25,-6.5) rectangle (1.75,-5.5);

\draw [black,-,dotted] (-4.25,-5.5) -- (-4.25,0.0) ;
\draw [black,-,dotted] (4.25,-5.5) -- (4.25,0.0) ;

\draw [black,->] (0,-6) -- (0,1.-6) node[left] {$\mathbf{\hat{z}}$};
\draw [black,->] (0,-6) -- (1.0,-6) node[above] {$\mathbf{\hat{x}}$};

\draw [black,->] (0.0,-6,0) -- (-2.65,-6.4,0.0) node[below] {$\bm{r}$};
\draw [black,->] (0.0,-6,0) -- (2.0,-6.3,0.0) node[below] {$\bm{r}_s$};

\draw [black,dotted] (2.0,-6.0,0) -- (2.0,-6.3,0.0) node[right] {$z_s$};
\draw [black,dotted] (-2.65,-6.0,0) -- (-2.65,-6.4,0.0) node[left] {$z$};
\draw [black,<->,dotted] (-2.65,-6,0) -- (2.0,-6,0.0);
\draw [black,<->,dotted] (2.0,-6.3,0) -- (-2.65,-6.4,0.0);
\draw [black,<->,dashed] (3.5,-6.5) -- (3.5,-5.5);
\node[text width=0.3cm] at (3.72,-6.0)  {$h$}; 

\node[text width=2cm] at (0.5,-5.9)  {$P$}; 
\node[text width=2cm] at (0.5,-6.3)  {$R$}; 

\draw [black,<->,dashed] (1.75,-5.3) -- (3.25,-5.3) ;
\node[text width=0.3cm] at (2.5,-5.1)  {$a$}; 

\end{tikzpicture}}
\caption{In-plane (top) and vertical (bottom) cross-sections of the quasi-planar geometry.}
\label{fig:scenario}
\end{figure}
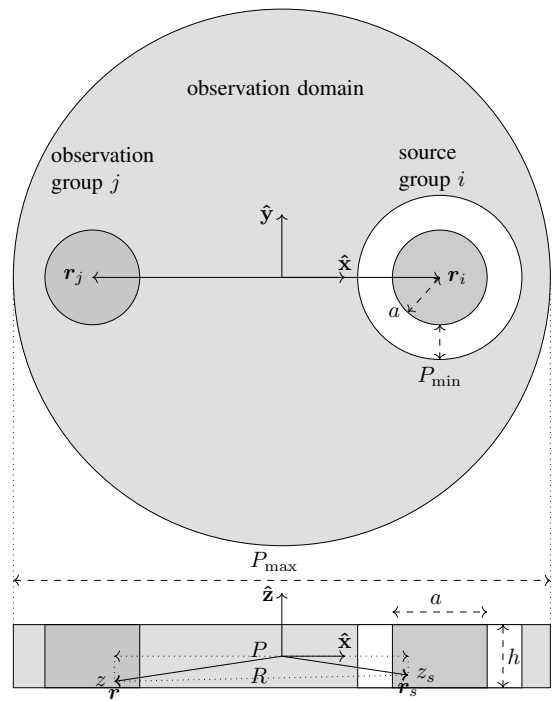

In this paper, we propose a broadband multipole-based decomposition that extends the approach in \cite{Bogaert2006} to address quasi-planar problems with small height over horizontal size ratio, as illustrated in Fig.~\ref{fig:scenario}. This is achieved by substituting the 2D multipole expansion into a line-source expansion of the 3D GF, akin to the high-frequency Steepest-Descent Fast Multipole Method (SDFMM, \cite{Jandhyala}). The Steepest-Descent Multipole (SDM) expansion thus includes the 2D multipole translation function and numerical integration along the wavenumber $k_z$, perpendicular to the array's horizontal plane along the Steepest Descent Path (SDP). At high frequencies, the SDM approach is more efficient for quasi-planar structures than standard 3D multipole expansions \cite{Rokhlin}, as it requires only plane waves propagating near the horizontal plane rather than across the entire sphere. At low frequencies, when antennas in dense wideband arrays are very close to each other, the required number of evanescent waves is drastically reduced by using a non-uniform sampling scheme along $k_z$, as devised in \cite{Chew1993,Gueuning}.
Since this broadband multipole method requires only pre-computation of complex patterns, i.e. far-field radiation patterns evaluated continuously both inside and outside the visible domain, for each MBF, the pre-processing step is cost-effective. This facilitates more efficient frequency sweeping and re-computation for various antenna geometries. Additionally, our method allows for the arbitrary rotation of antennas at low additional computational cost. We demonstrate that full simulations with arrays of 256 wideband log-periodic antennas can be completed in just 10 minutes on a laptop, providing a significant performance improvement compared to FEKO's MLFMM \cite{FEKO}, which is also constrained by the LF breakdown, and to HARP\cite{HARP}.

The paper is organized as follows. We start by defining the geometrical quantities for the quasi-planar structure in Section~\ref{sec:geometry}.
Section~\ref{sec:HFmultipole} recalls the SDM expansion of the free-space GF. In Section~\ref{sec:numericalissues}, we then identify the two numerical issues appearing at subwavelength distances. The broadband expansion resolving these issues is presented in Section~\ref{sec:broadbandmultipole}. This expansion is then used to accelerate the MoM solver described in Section~\ref{sec:MoM}. Numerical examples are presented in Section~\ref{sec:numexamples}. Finally, conclusions are drawn in Section~\ref{sec:Conclusion}.

\section{Quasi-planar geometry}
\label{sec:geometry}
We consider a quasi-planar structure confined within a cylinder of height $h$ and diameter $P_{\max}$, as shown in Fig.~\ref{fig:scenario}.
We define a cylindrical system of coordinates $(\hat{\bm{\rho}}, \hat{\bm{\phi}},\hat{\mathbf{z}})$ and a Cartesian system of coordinates $(\hat{\bm{x}}, \hat{\bm{y}},\hat{\bm{z}})$. A source point $\mathbf{r}_s = x_s \hat{\bm{x}}+ y_s\hat{\bm{y}}+z_s\hat{\mathbf{z}}$ is represented by the cylindrical coordinates $(\rho_s, \phi_s, z_s)$, while an observation point $\mathbf{r}= x \hat{\bm{x}}+ y\hat{\bm{y}}+z\hat{\mathbf{z}}$ is denoted by coordinates $(\rho, \phi, z)$. The distance between these points is given by $R = (P^2+(z-z_s)^2)^{1/2}$ where the in-plane distance is defined by $P = ((x-x_s)^2+(y-y_s)^2)^{1/2}$. A group $i$ is enclosed within a cylinder of height $h$, radius $a$ and center $\mathbf{r}_i =  x_i \hat{\bm{x}}+ y_i\hat{\bm{y}}$ which lies in the $z = 0 $ plane. The in-plane distance $P$ between any pairs of source and observation points ranges from $P_{\min}$ to $P_{\max}$.

\section{Reminder on the Steepest-Descent Multipole expansion of the three-dimensional free-space Green's function}
\label{sec:HFmultipole}
\begin{figure}[h]
\centering
\resizebox {0.45\textwidth} {!} {
\begin{tikzpicture}
\draw[->] (-5,0) -- (5,0) node[below=0.2cm] {\large $k_{z r}$};
\draw[black,] (1.5,-0.1) -- (1.5,0.1) node[below=0.2cm] {\large $k$};		
\draw[black,] (-1.5,-0.1) -- (-1.5,0.1) node[below=0.2cm] {\large $-k$};		
\draw[black] (-0.1,2) -- (0.1,2) node[left=0.2cm] {\large $ k$};
\draw[black] (-0.1,-2) -- (0.1,-2) node[left=0.2cm] {\large $- k$};
\draw[->] (0,-3) -- (0,3) node[anchor=east] {\large $k_{z i}$};

\draw[scale=1,domain=-5:5,smooth,variable=\x,black,very thick] plot ({\x},{ \x / ( 1.0 + (\x)^2.0/4.0)^0.5 });
\end{tikzpicture}}
\caption{Illustration of the Steepest Descent Path (SDP).}
\label{fig:contours}
\end{figure}
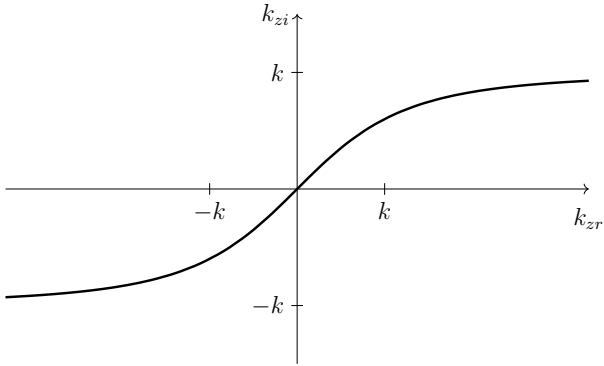

We begin with a reminder of the Steepest-Descent Multipole (SDM) decomposition \cite{Jandhyala}. This starts with the decomposition of the free-space 3D Green's function (GF) into a spectrum of 2D line sources as follows:
\begin{align} \nonumber
G(k,R) & = \frac{e^{-jkR}}{4\pi R} \\ 
& = \frac{-j}{8\pi} \int\limits_{-\infty}^{\infty}  H_0^{(2)}(k_\rho P) \ e^{ -jk_{z} (z-z_s)} \  \mathrm{d}k_{z}
\label{Greenfunction}
\end{align}
where $k$ is the wavenumber,  $k_z$ and $k_\rho = (k_x^2+k_y^2)^{1/2} = (k^2-k_z^2)^{1/2}$ are the vertical and radial spectral components of the wavevector $\mathbf{k} = k_x \hat{\bm{x}}+ k_y\hat{\bm{y}}+k_z\hat{\mathbf{z}}$ and $H_0^{(2)}$ is the second-kind Hankel function of zero order. A contour deformation in the complex plane must then be used to avoid numerical issues stemming from the singularity of the Hankel function which appears for $k_{zr} = k$. Since the function $H_0^{(2)}(k_{\rho}P)$ oscillates as $e^{-jk_{\rho r}P}$, it can be efficiently integrated along the Steepest Descent Path (SDP,\cite{SDP}) where $k_{\rho,r}$ is kept constant, thus removing the modulation. The SDP is illustrated in Fig.~\ref{fig:contours} and is defined by the following equations:
\begin{align}
k_{zi} = \frac{k_{zr}}{S}
\label{SDP}
\end{align}
where $S = (1+ (k_{zr}/k)^2)^{1/2}$ and the contour derivative is given by $k_{zi}' = 1/S$.
It must be noted that the SDP is not suited to structures with height $h$ much larger than the wavelength because the amplitude of the Hankel function grows significantly in the visible region $k_{zr} \in [-k, k]$ causing large round-off errors \cite{Gueuning4}. In such cases, a contour with a lower maximum height $k_{zi}$ is preferable.

The next step involves expanding the Hankel function in \eqref{Greenfunction} using Rokhlin's 2D multipole expansion \cite{Rokhlin}. For an observation point $\boldsymbol{\mathbf{r}}$ in the group $i$ and a source point $\boldsymbol{\mathbf{r}}_s$ in the group $j$, we can write
\begin{align} \nonumber
H_0^{(2)}(k_\rho P) = \frac{1}{2\pi} \int\limits_{0}^{2\pi} e^{j\mathbf{k}_\rho \cdot (\boldsymbol{\mathbf{r}}-\mathbf{r}_
j)} \  T(k P_{ij},\alpha-\phi_{ij})  \\  e^{ -j\mathbf{k}_\rho \cdot (\boldsymbol{\mathbf{r}}_s-\mathbf{r}_i)} \  \mathrm{d}\alpha
\label{Rokhlin2d}
\end{align}
where $P_{ij}$ and $\phi_{ij}$ are the polar coordinates of the in-plane relative distance $\boldsymbol{P}_{ij} = \mathbf{r}_j-\mathbf{r}_i$, $\alpha$ represents the azimuthal angle associated with a plane wave with wavector $\mathbf{k}$, and its horizontal projection is $\mathbf{k}_\rho = k_{\rho}\cos\alpha \ \hat{\mathbf{x}} + k_{\rho}\sin\alpha \ \hat{\mathbf{y}}$. The translation function of the 2D multipole expansion is written as
\begin{align}
T(k_\rho P_{ij},\alpha) =  \sum_{m=-\infty}^{\infty}  j^{m}  \  H_{m}^{(2)} (k_\rho P_{ij}) \ e^{jm \alpha}
\label{translation0}
\end{align}
where $H_{m}^{(2)}$ is the second-kind Hankel function of order $m$. Finally, susbtituting \eqref{Rokhlin2d} into the line-source decomposition \eqref{Greenfunction} leads to the SDM decomposition of the 3D GF \cite{Jandhyala},
\begin{align} \nonumber
G(k,R) = \frac{-j}{16\pi^2}  \int\limits_{-\infty}^{\infty}  \int\limits_{0}^{2\pi}  & e^{j\mathbf{k} \cdot (\mathbf{r}-\mathbf{r}_j)} \  T(k_\rho P_{ij},\alpha-\phi_{ij})  \\ & e^{ -j\mathbf{k} \cdot (\mathbf{r}_s-\mathbf{r}_i)} \  \mathrm{d}\alpha \  k_z' \mathrm{d}k_{zr}
\label{planewavedecomp}
\end{align}
where $k_z' = (1+j k_{zi}')$ accounts for the contour derivative. The expression \eqref{planewavedecomp} differs from the 3D multipole decomposition \cite{Rokhlin} as it uses the 2D translation function and integration along the vertical wavenumber $k_z$. For quasi-planar geometries, it only requires plane waves propagating near the horizontal plane, i.e. at small $k_{zr}$ values. The number of samples along $k_{zr}$ also scales nearly linearly with the horizontal electrical size $ka$ \cite{Gueuning,Jandhyala}, unlike the quadratic scaling $(ka)^2$ in traditional 3D multipole expansion. Sampling rules for discretizing \eqref{planewavedecomp} at high frequencies are provided in \cite{Gueuning}.

\section{Numerical challenges at sub-wavelength distances}
\label{sec:numericalissues}
For small relative distances $P_{ij}$ below half a wavelength, the multipole decomposition \eqref{Rokhlin2d} becomes inaccurate and the required number of samples along $k_{zr}$ rapidly increases due to the need for more evanescent waves. In this section, we delve deeper into these two numerical challenges. In the next section, we will present the techniques that lead to the broadband extension of the SDM expansion.

\subsection{Low-frequency (LF) breakdown of the multipole expansion}
The LF breakdown of the multipole decomposition is a well-known numerical issue \cite{Chew1999}, \cite{Bogaert2008} that arises when the translation function in \eqref{Rokhlin2d} is evaluated at a small argument $kP_{ij}$, i.e., at low frequencies and/or small distances. The ill-conditioning results from an attempt to evaluate reactive near fields, with spatial spectrum extending far beyond the visible domain, using only far-field information. Mathematically, the breakdown is attributed to the over-exponential asymptotic growth of the Hankel functions $H_m^{(2)}$ w.r.t. order $m$ in \eqref{translation0} for small arguments $|k_\rho P_{ij}|<m$ \cite{Abramowitz}
\begin{align}
H_{m}^{(2)} (k_\rho P_{ij}) \approx \frac{-j}{\pi} \ (m-1)! \ 2^m \  (k_\rho P_{ij})^{-m}
\label{Hmasymptotic}
\end{align}
where $(m-1)!$ is the asymptotically dominant function. Despite their similar importance to overall accuracy, the low-order terms in \eqref{translation0} have much smaller amplitudes compared to the higher-order ones. For instance, when $M=10$, $k_\rho = k$, and $P_{ij} = 0.1 \lambda$, we have $|H_M^{(2)} (k_\rho P_{ij}) / H_0^{(2)} (k_\rho P_{ij})| \approx 10^{10}$. Due to finite machine precision, the amplitude of low-order terms can be smaller than the round-off noise of higher-order terms, leading to numerical breakdown.

\subsection{Wide evanescent spectrum}
\label{subsec:issue2}
The second numerical issue is related to the computation of the outer integral along $k_{zr}$. For simplicity, we can focus solely on the line-source decomposition \eqref{Greenfunction} and consider the in-plane case $z-z_s = 0$. The integrand is then decaying exponentially at large $k_{zr}$ values for which $k_\rho \approx jk_{zi} $.  Using the large-argument approximation of $H_0^{(2)}(x)$ \cite{Abramowitz}, the relative truncation error is estimated as:
\begin{align}
\epsilon & \approx  \sqrt{2P/\pi}  \int_{k_{z m}}^\infty  \frac{e^{-k_{zr} P}}{\sqrt{k_{zr}}}  \  \mathrm{d}k_{z r} \nonumber \\
 		& = \sqrt{2} \ \erfc(\sqrt{k_{zm} P})
\label{truncerr}
\end{align}
where $\erfc$ is the complementary error function and $k_{zm}$ is the truncation limit.
Using $\erfc(x)<e^{-x^2}$, we obtain a bound for $k_{zm}$,
\begin{align}
k_{zm} \gtrsim \frac{-\ln(\epsilon/\sqrt{2})}{P}
\label{truncrhozero}
\end{align}
On one hand, the truncation limit $k_{zm}$ must  increase as the distance $P$ decreases to account for the more singular behavior of the GF by including higher spatial frequencies.
For instance, for a target error level of $\epsilon = 10^{-2}$ and $P = 0.01\lambda$, the integration must be performed up to $k_{zm} \sim 80k$, far beyond the visible limit which is located at $k_{zr} = k$. On the other hand, the sampling step $\Delta k$ used to discretise \eqref{Greenfunction} must decrease linearly with increasing distance $P$ as result of the Gaussian decay of $H_0^{(2)}$ along the SDP at low $k_{zr}$.
To devise an efficient expansion, the evaluation of \eqref{Greenfunction} with a given $\Delta k$ and $k_{zm}$ must remain accurate for a range of distances $P$ covering the entire observation domain, i.e., with $P \in [P_{\min}, P_{\max}]$. For instance, when $P_{\min} = 0.01 \lambda$ and $P_{\max} = 10 \lambda$, this approach would require thousands of samples along $k_{zr}$.

\section{Broadband Steepest-Descent Multipole formulation}
\label{sec:broadbandmultipole}
We now present a broadband SDM decomposition of the GF, addressing the two numerical issues previously described. Firstly, we resolve the LF breakdown of the 2D multipole decomposition by employing a re-normalization approach similar to \cite{Chew1999, Bogaert2006,Meng}, but we retain the plane-wave formulation instead of using azimuthal harmonics. Secondly, we tackle the challenge of the extensive evanescent spectrum through non-uniform sampling. Finally, we validate the multipole expansion of the 3D GF function with numerical experiments.

\subsection{Stabilitization of the multipole expansion}
\label{subsec:stab}
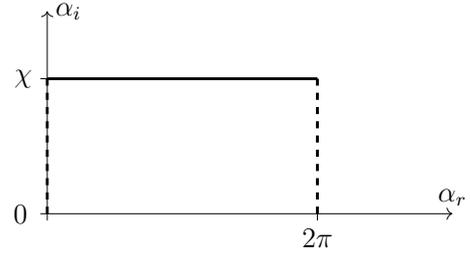
\begin{figure}[htbp]
\centering
\resizebox {0.35\textwidth} {!} {
\begin{tikzpicture}
\draw[->] (-0.1,0) -- (6,0) node[above] {\large $\alpha _r$};	
\draw[ black] (4,-0.1) -- (4,0.1) node[below=0.2cm] {\large $2\pi$};	

\draw[black] (-0.1,2) -- (0.1,2) node[left=0.2cm] {\large $ \chi $};
\node[text width=1cm] at (0.0,-0.0) 
    {\large $0$}; 

\draw[->] (0,-0.1) -- (0,3) node[anchor=west] {\large $\alpha _i$};

\draw[very thick,black,dashed] (0,0) -- (0,2);
\draw[very thick,black,dashed] (4,0) -- (4,2);

\draw[scale=1,domain=0:4,variable=\x,black,very thick] plot ({\x},{ 2});
\end{tikzpicture}}
\caption{Illustration of a path shifted by a constant $\chi$ in the complex azimuthal plane.}
\label{fig:contourshift}
\end{figure}

Let us first define the one-sided translation function, which includes only the positive orders of the translation function $T$, as follows:
\begin{align}
T_o(kP_{ij},\alpha) =  \sum_{m=0}^{\infty}  c_m \ j^{m}   \  H_{m}^{(2)} (kP_{ij})  \   e^{ jm \alpha}
\label{onesidetranslation}
\end{align}
where  $c_m = 1/2$ for $m=0$ and $c_m = 1$ otherwise. Observing the relation $T(kP_{ij},\alpha) = T_o(kP_{ij},\alpha) + T_o(kP_{ij},-\alpha)$, we can then rewrite the decomposition \eqref{Rokhlin2d} of the Hankel function into two terms, 
\begin{align}
H_0^{(2)}(k P) = H_+ (kP) + H_- (kP)
\label{pmHankel}
\end{align}
where $H_\pm$ are functions obtained by evaluating the multipole decomposition with the one-sided translation function $T_o$:
\begin{align}\nonumber
H_\pm(k P) = \int\limits_{0}^{2\pi} e^{j\mathbf{k}_\rho \cdot (\boldsymbol{\mathbf{\rho}}-\boldsymbol{\mathbf{r}}_j)} \ & T_o(k P_{ij},\pm (\alpha-\phi_{ij}))  \\   & e^{ -j\mathbf{k}_\rho \cdot (\boldsymbol{\mathbf{\rho}}_s-\boldsymbol{\mathbf{r}}_i)} \ \mathrm{d}\alpha
\label{Hpm}
\end{align}
This splitting over the indices $m$ of the translation function $T$ is equivalent to that performed over the indices $k$ in the equations (12) and (14) in \cite{Bogaert2006}.
As outlined in \cite{Bogaert2008}, we deform the integration path along the azimuthal angle $\alpha$ to a contour in the complex plane with three segments: one horizontal segment parallel to the real axis with shift $\alpha_i = \pm \chi$ for $H_{\pm}$ respectively, and two vertical segments parallel to the imaginary axis at $\alpha_r = 0$ and $\alpha_r = 2\pi$, as depicted in Fig.~\ref{fig:contourshift} for $H_+$. Since the integrand is $2\pi$-periodic, the contributions along the two vertical branches cancel out, leaving only the horizontal segment. Evaluating the one-sided translation function at $\alpha = \pm \alpha_r + j\chi$ produces a decreasing exponential factor $e^{-m \chi}$ as a function of order $m$,
\begin{align}
T_o(kP,\alpha) =  \sum_{m=0}^{\infty}  c_m \ j^{m}   \  H_{m}^{(2)} (kP) \  e^{ -m \chi} \ e^{\pm jm \alpha_r}
\label{translation3}
\end{align}
The exponential factor $e^{ -m \chi}$ counteracts the factorial increase \eqref{Hmasymptotic} of  $H_m^{(2)}$, thereby re-normalising the amplitude of each term. When the sum in \eqref{translation3} is truncated to a maximum order $M$, the complex shift $\chi$ can be chosen such that the ratio between the highest-order ($m=M$) and the lowest-order ($m=0$) Hankel functions in the translation function is balanced by the exponential,
\begin{align}
\chi(k) = \frac{1}{M} \log \frac{|H_M^{(2)}(kP_{\min})|}{|H_0^{(2)}(k P_{\min})|}
\label{chi}
\end{align}
where the dependency of $\chi$ on $k$ is introduced, as we will simply replace $k$ with the complex value $k_\rho$ when using this formula for the 3D expansion.

\subsection{Non-uniform sampling along the vertical wavenumber $k_{zr}$}
\begin{figure}[]
\subfloat[]{\includegraphics[trim=0.0cm 0.0cm 0.0cm 0.5cm,clip,width=9cm,height=4cm]{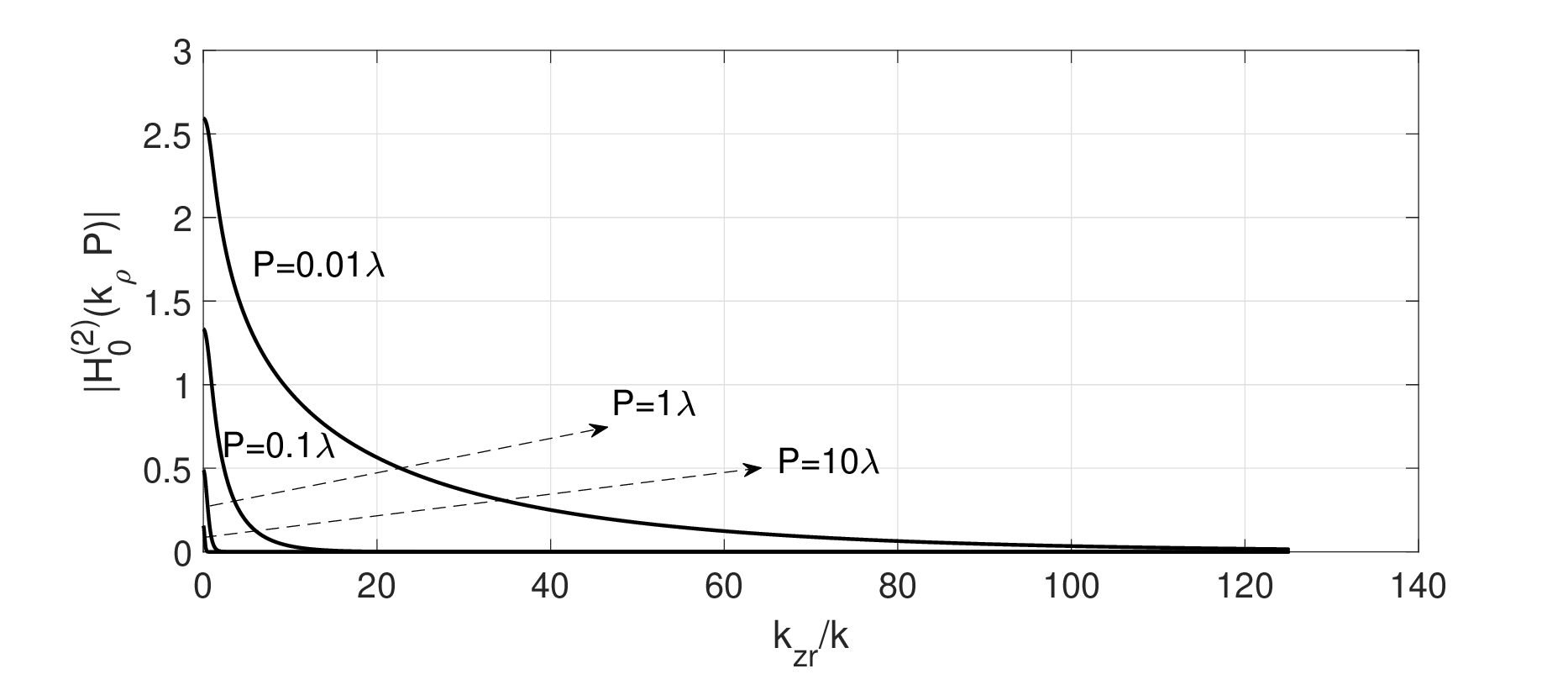}}\\
\subfloat[]{\includegraphics[trim=0.0cm 0.0cm 0.0cm 0.50cm,clip,width=9cm,height=4cm]{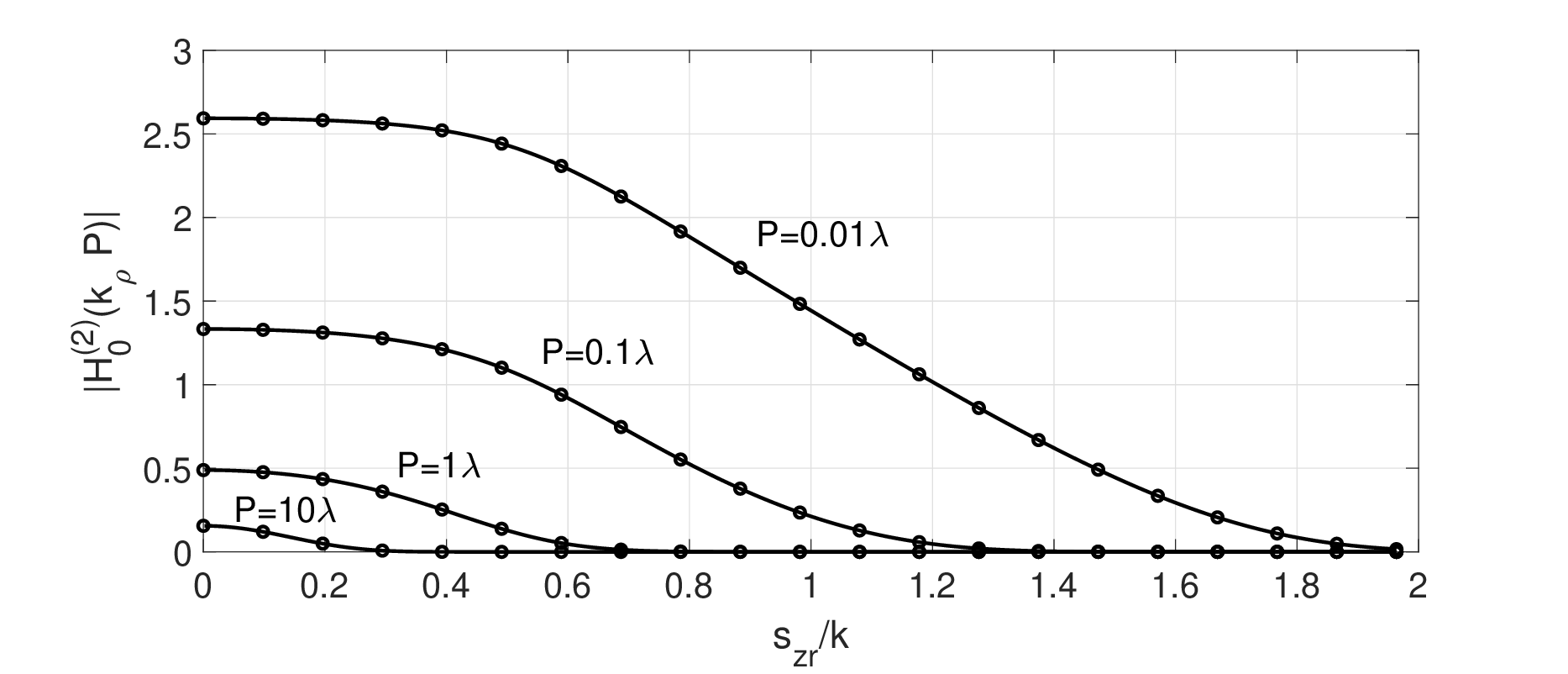}}\\
\subfloat[]{\includegraphics[trim=0.0cm 0.0cm 0.0cm 0.50cm,clip,width=9cm,height=4cm]{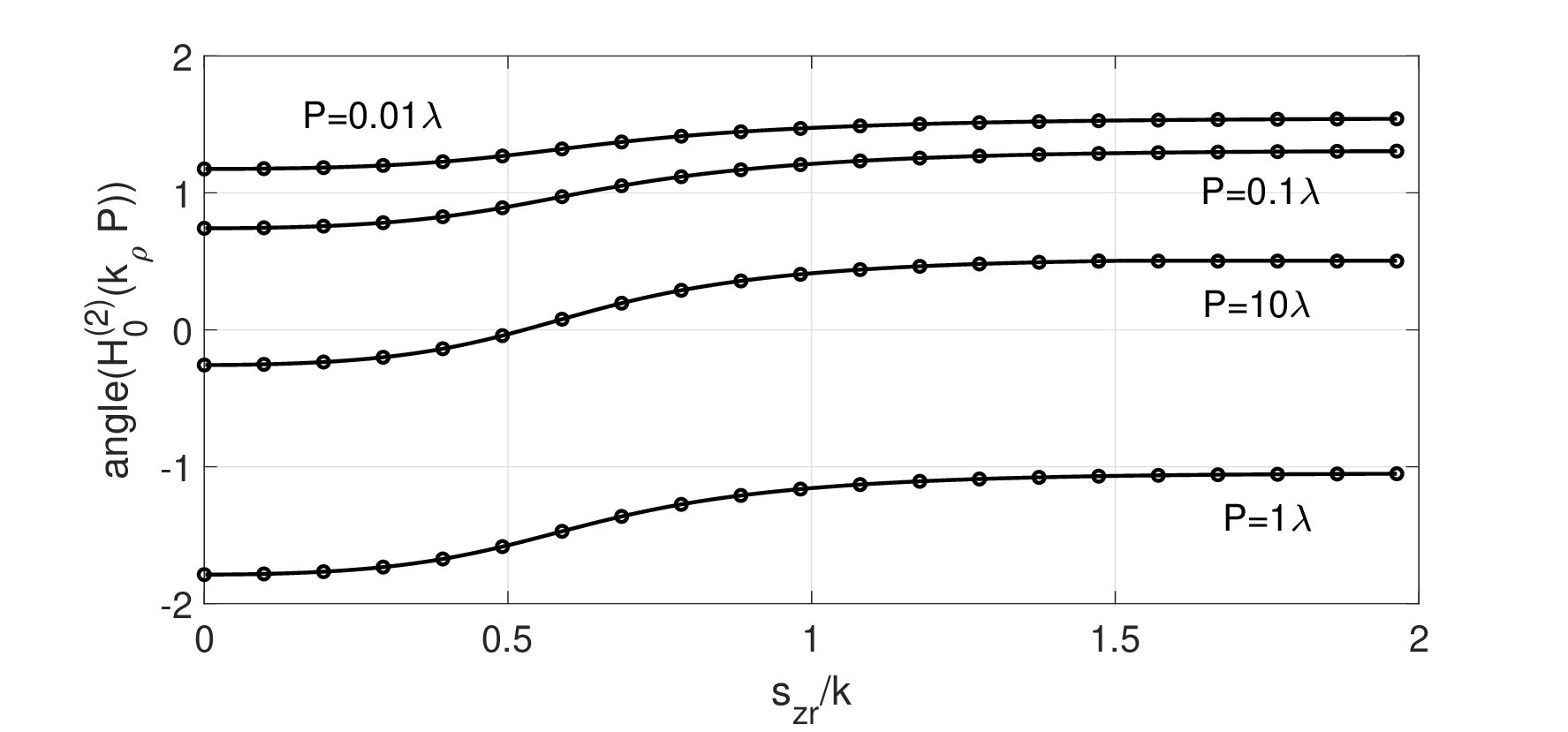}}
\caption{Amplitude and phase of the Hankel function $H_0^{(2)}$ along the SDP without (a) and with (b), (c) the change of variable \eqref{changevar} at distances $P = \{0.01,0.1,1,10\}\lambda$. For a target error $\epsilon = 10^{-4}$, we have $k_{zm}/k = 125$,  $\Delta s / k = 0.1$,  $b = 1.31$ and $N_z = 21$.}
\label{fig:changevar}
\end{figure}

As discussed in Section~\ref{subsec:issue2}, the required number of plane waves with uniform sampling along $k_{zr}$ becomes prohibitively high for observation domains with small distances $P_{\min}$ because the truncation limit $k_{zm}$ increases substantially while the step size $\Delta k$ is still constrained by the variation of $H_0^{(2)}$ at the largest distance $P_{\max}$. To address this, we adopt a non-uniform sampling along $k_{zr}$ with a step size $\Delta k$ that increases linearly with $k_{zr}$. This is motivated by the observation \cite{Chew1993} that plane waves with larger $k_{zr}$ values contribute significantly only at smaller distances $P$, for which $H_0^{(2)}$ is decaying at a slower rate. According to \cite{Gueuning}, the non-uniform sampling can be implemented using the following change of variables:
\begin{align}
s_{zr} = \asinh(b k_{zr}) / b
\label{changevar}
\end{align} 
where $\asinh$ is the inverse hyperbolic sine, $b$ is a parameter and $\mathrm{d}k_{zr}/\mathrm{d}s_{zr} = \cosh (b s_{zr})$ is the Jacobian of the transformation. When $bk_{zr} \gg 1$, the sampling along $k_{zr}$ is almost regular since we have $s_{zr} \simeq k_{zr}$. For larger $s_{zr}$, the sampling increases exponentially \cite{Chew1993}, with $k_{zr} \simeq 1/(2b) e^{bs_{zr}}$. As highlighted in \eqref{truncerr}, at large $k_{zr}$, the Hankel function behaves as a decaying exponential. Therefore, the exponential sampling is designed to counteract this exponential decay,
\begin{align}
e^{-k_{zr} P} \sim  e^{-P/(2b) e^{bs_{zr}}}
\label{changevar2}
\end{align}
As illustrated in Fig.~\ref{fig:changevar}, the change of variables not only smoothens out the function  $H_0^{(2)}$ w.r.t. $k_{zr}$ but also reduces its variation w.r.t. $P$, allowing integration with fewer points. 

Following \cite{Gueuning}, the new variable $s_{zr}$ can be sampled on a grid with spacing $\Delta s$ and truncation limit $s_{zm}$ given by
\begin{align} \nonumber
\Delta s & = \pi (- k/(P_{\max}\ln\epsilon)^{1/2} \\
s_{zm} & = \asinh(b k_{zm}) / b
\label{changevar2}
\end{align}
For purely planar cases ($h=0$), we can select $b = \pi / (-\ln \epsilon \Delta s)$ and $k_{zm}$ with \eqref{truncrhozero}. The number of samples $N_z$ now scales with \cite{Gueuning}, 
\begin{align} 
N_z = \frac{2s_{zm}}{\Delta s}  \propto -\ln \epsilon \ \ln \frac{P_{\max}}{P_{\min}}
\label{Nz}
\end{align}
instead of being inversely proportional to $P_{\min}$ as with regular sampling.
For $\epsilon = 10^{-4}$, we have $k_{zm}/k = 125$,  $\Delta s / k = 0.1$,  $b = 1.31$, this approach only requires $N_z = 21$ samples. 
For non-planar structure ($h>0$), the oscillation with $z-z_s$ in \eqref{Greenfunction} must be considered. To prevent the sampling from exceeding the Nyquist rate $\pi/h$ near the end of the integration of the integration domain, a slightly lower value of $b$ can be chosen with the following rule of thumb:
\begin{align}
b \sim \min(\frac{\pi}{-\ln \epsilon \Delta s}, \frac{\gamma \pi}{k_{zm} h\Delta s})
\label{changevar2}
\end{align}
where $\gamma \simeq 1$ is manually adjusted. The truncation limit $k_{zm}$ can be selected with \eqref{truncrhozero} but the target error level $\epsilon$ must be multiplied with $e^{hk}$ to account for the exponential increase $e^{k_{zi}(z-z_s)}$ of the integrand in \eqref{Greenfunction} along the SDP.

\subsection{Numerical evaluation of the 3D GF}
\label{subsec:3DGF}
We can now analyse the 3D problem by substituting the expansion \eqref{pmHankel} of $H_0^{(2)}$ into \eqref{Greenfunction}, along with the transformation \eqref{changevar} and the two contour deformations. Discretizing both integrals leads to:
\begin{align}
G(k,R) = G_+ (k,R) + G_- (k,R)
\end{align}
with
\begin{align}\nonumber
G_\pm (k,R) =  \sum_{p=1}^{N_s} \sum_{n=0}^{2M}  w_p  \ &  T_o(k_{\rho,p} P_{ij},\pm(\alpha_{r,n}-\phi_{ij})+j \chi_p)  \\ & e^{j\mathbf{k}_{pn} \cdot (\mathbf{r}-\mathbf{r}_j)} \  e^{ -j\mathbf{k}_{pn} \cdot (\mathbf{r}_s-\mathbf{r}_i)}
\label{num3DGF}
\end{align}
where the wavevector of the plane wave $pn$ is $\mathbf{k}_{pn} = k_{\rho,p} \cos \alpha_n \hat{\mathbf{x}} + k_{\rho,p} \sin \alpha_n \hat{\mathbf{y}} + k_{z,p} \hat{\mathbf{z}}$, the integration weight is $w_p = k_{z,p}'  \cosh (b s_{zr,p}) \Delta s$ and the contour shift is $\chi_p = \chi(k_{\rho,p})$ according to \eqref{chi}. 

\begin{figure}[h]
\centering
\subfloat[]{\includegraphics[trim=0.0cm 0.0cm 0.0cm 1.0cm,clip,width=8.5cm,height=6.5cm]{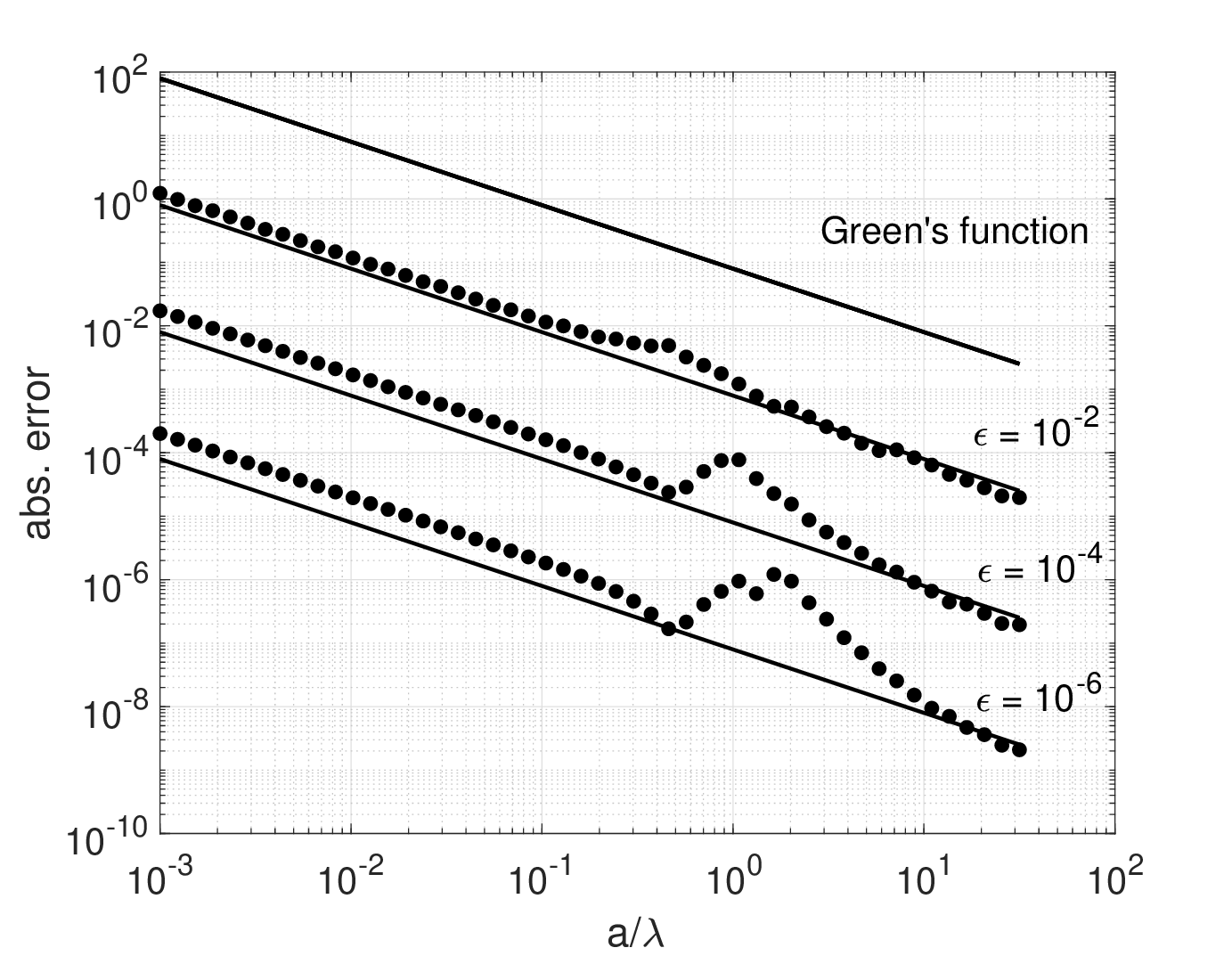}}\\
\subfloat[]{\includegraphics[trim=0.0cm 0.0cm 0.0cm 1.0cm,clip,width=8.5cm,height=6.5cm]{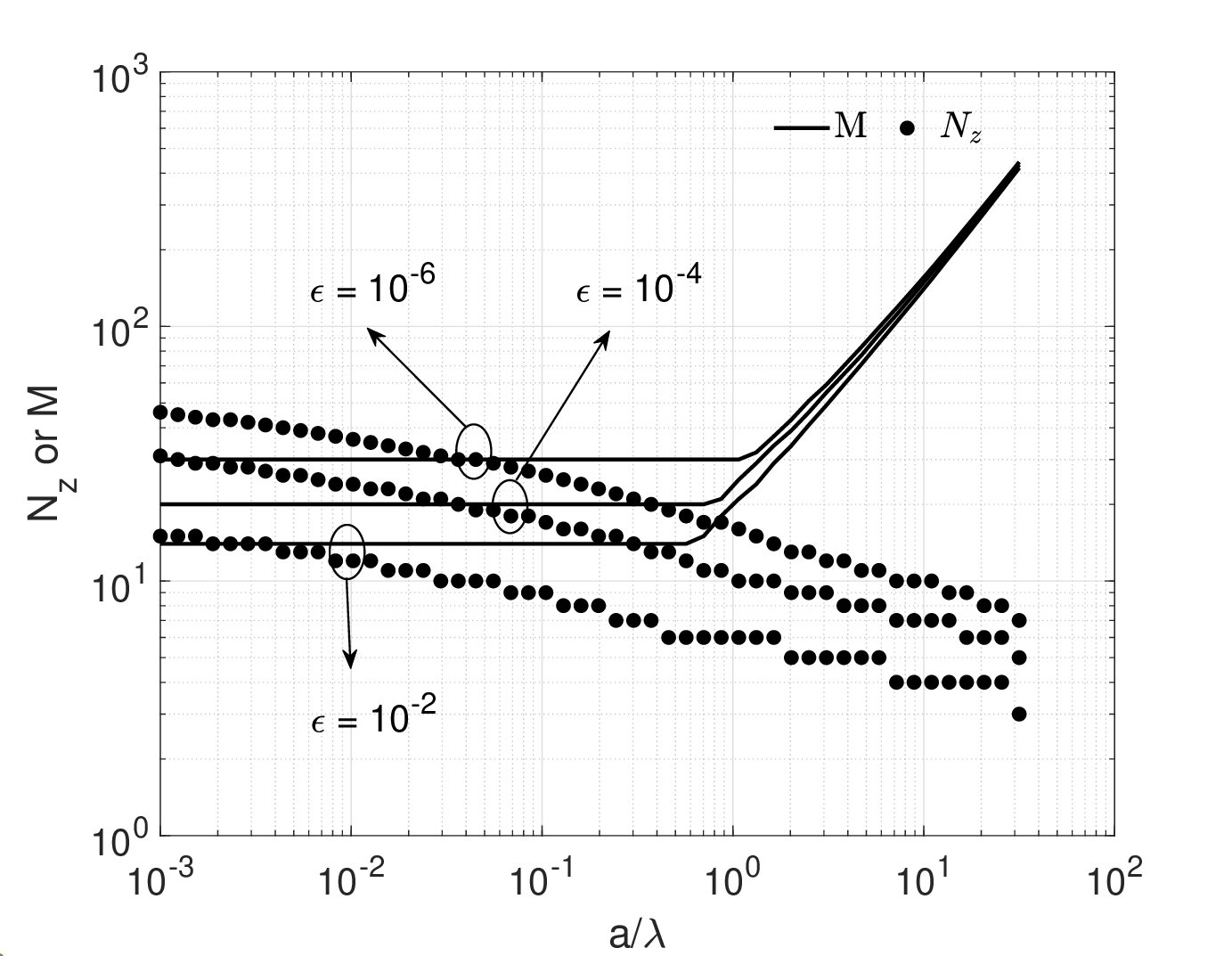}}
\caption{Numerical evaluation of the 3D GF for a planar scenario ($h = 0$) with group size $a/\lambda$, minimum distance  $P_{\min} = a$, $P_{\max} = 50\lambda$. The maximum absolute errors (a) and the number of samples $N_z$ and $M$ (b) are shown for $\epsilon = \{10^{-2}, 10^{-4}, 10^{-6}\}$ and increasing size $a$.}
\label{fig:validationGF}
\end{figure}

First, we validate the numerical evaluation of the GF through an purely planar experiment ($h=0$) where the source group size $a$ increases from electrically small ($a=0.001\lambda$) to electrically large ($a = 25\lambda$). The minimum distance between source and observation domains increases with the group size as $P_{\min} = a$. The maximum distance is fixed at $P_{\max} = 50\lambda$. As shown in Fig.~\ref{fig:validationGF}.a, we achieve error levels $\epsilon = \{10^{-2}, 10^{-4}, 10^{-6}\}$ across the full frequency range, confirming that our representation is broadband and stable at very low frequencies. The truncation order $M$ of the translation function and the number $N_z$ of samples along $k_{zr}$ are analyzed in Fig.~\ref{fig:validationGF}.b. The order $M$ is kept constant in the LF regime and progressively converges to the excess bandwidth formula \cite{Chewbook}, i.e. increasing linearly with $a/\lambda$. The number $N_z$ increases with $-\ln P_{\min}$, as predicted. The bump around $a/\lambda = 1$, in the intermediate-frequency range of the error curves, is simply due to the transition from the LF sampling rules to the HF rules derived in \cite{Gueuning} and can be accommodated by choosing a slightly higher target error level.

\begin{figure}[h]
\centering
\subfloat[]{\includegraphics[trim=0.0cm 0.0cm 0.0cm 1.0cm,clip,width=8.5cm,height=6.5cm]{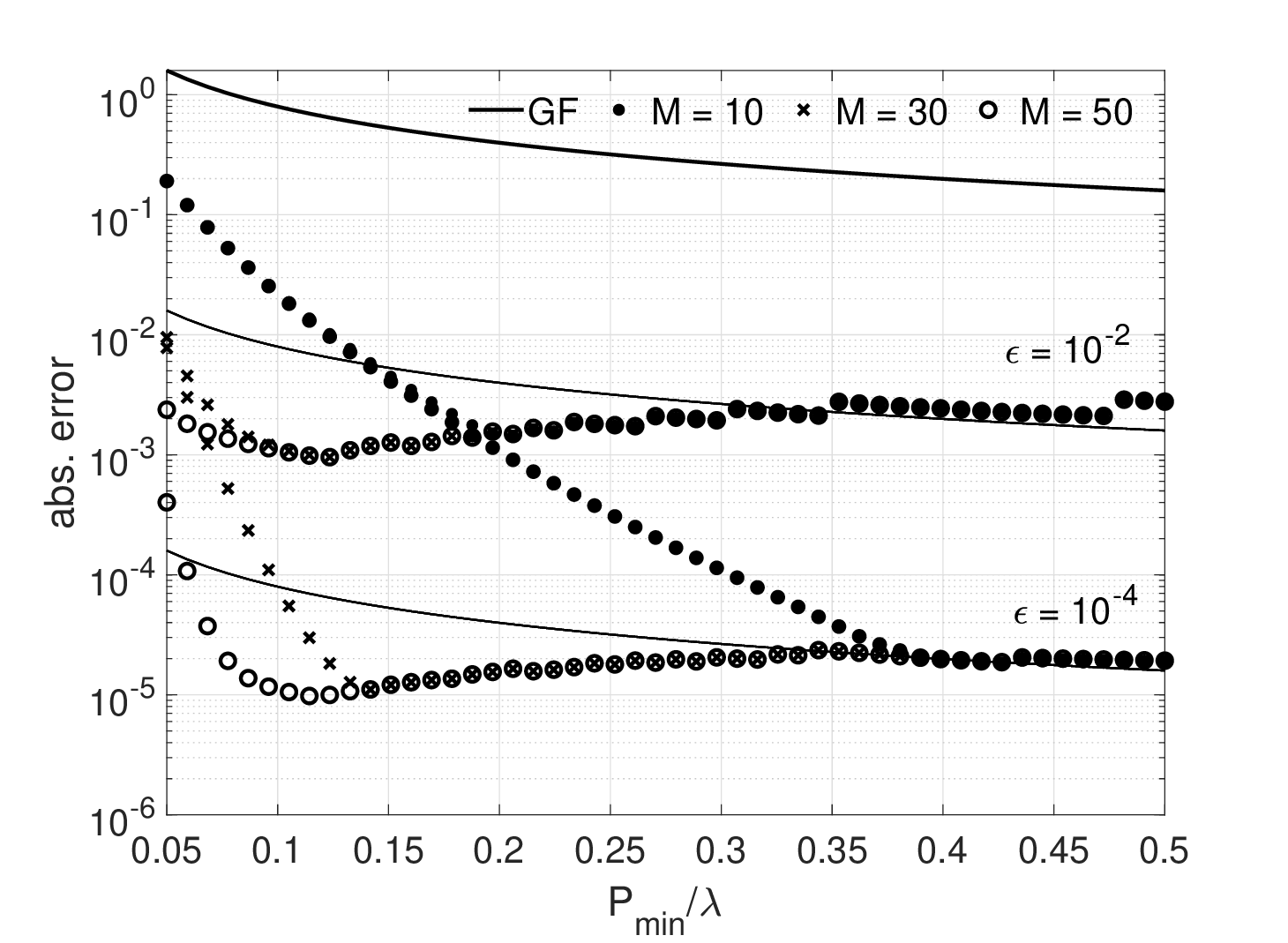}}\\
\subfloat[]{\includegraphics[trim=0.0cm 0.0cm 0.0cm 0.5cm,clip,width=8.5cm,height=4.2cm]{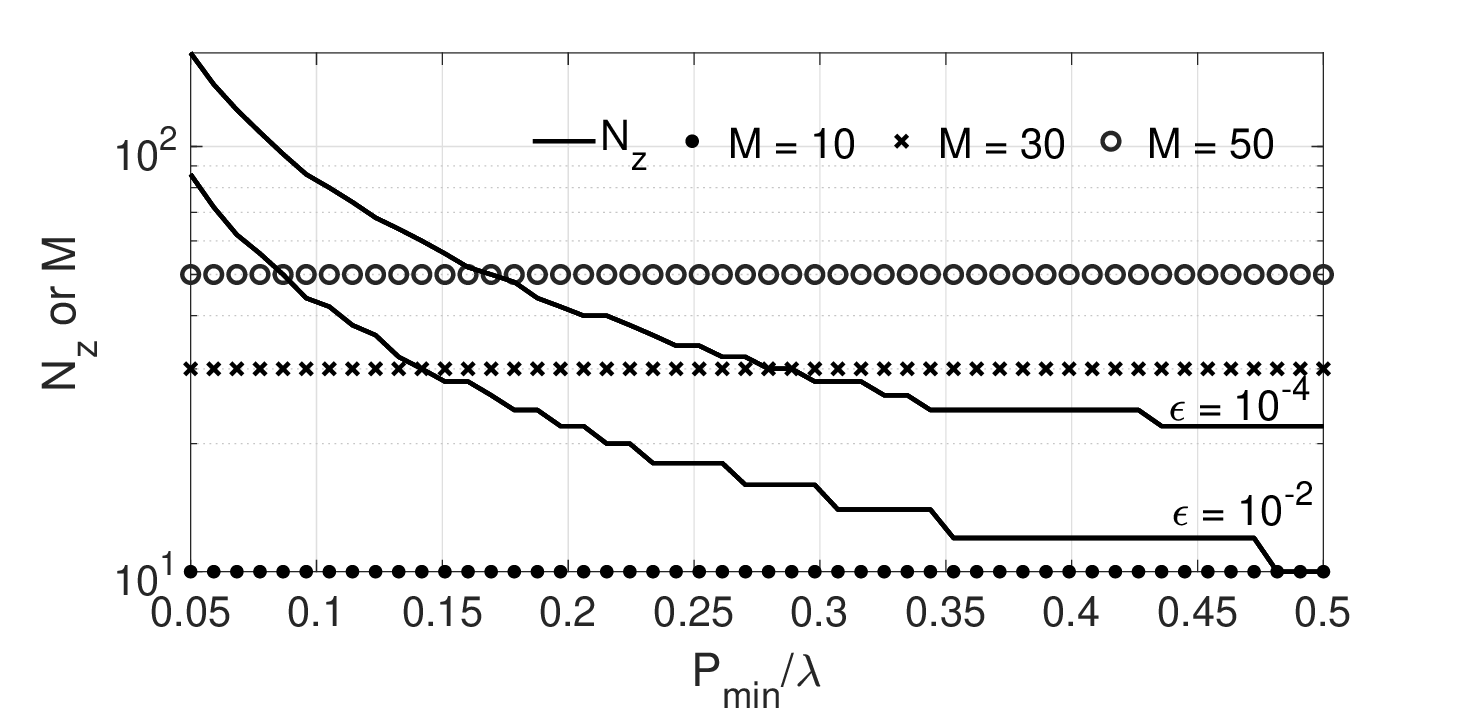}}
\caption{Numerical evaluation of the 3D GF for a non-planar scenario ($h = \lambda/3$) with fixed group size $a = \lambda/6$ and $P_{\max} = 50\lambda$. The minimum distance is varying between $P_{\min} = [0.05, 0.5]\lambda$. The maximum absolute errors (a) and the number of samples $N_z$ and $M$ (b) are shown for $\epsilon = \{10^{-2}, 10^{-4}\}$.}
\label{fig:validationGF2}
\end{figure}

Second, we propose a non-planar experiment and analyse the variation of the maximum error with $N_z$ and $M$ at the minimum distance $P_{\min}$. To be representative of typical 3D antenna sizes, we consider a cylinder of fixed height $h=\lambda /3$ and group size $a = \lambda/6$. The maximum distance is set to $P_{\max} = 50\lambda$. The minimum distance ranges $P_{\min}$ from $0.5 \lambda$ to $0.05 \lambda$. As shown in Fig.~\ref{fig:validationGF2}.a, increasing the multipole order $M$ help decrease the error at the smallest distance, with $M=30$ leading to an error of $10^{-2}$ and $M=50$ to an error of $10^{-4}$. One can see in Fig.~\ref{fig:validationGF2}.b that, for these minimum distances $P_{\min}$, the number of samples $N_z$ remains reasonable between $10$ and $100$. The minimum distance of $0.05 \lambda$ considered here is an order of magnitude lower than where the classical multipole expansion breaks down, which is around $0.5 \lambda$, as shown later in Fig.~\ref{fig:MultipoleInteraction}. 

\section{Accelerated Method-of-Moments solver}
In this section, we apply the SDM decomposition for the rapid evaluation of mutual Method-of-Moments (MoM) interactions. We begin with a brief overview of the MoM system of equations and the Macro-Basis Function (MBF) framework. Then, we introduce the SDM formulation of the MoM interactions and demonstrate how to re-compute these interactions for rotated antennas. Finally, we validate the method with a numerical test involving two log-periodic antennas.

\label{sec:MoM}
\subsection{Reminder on the MoM-MBF framework}
Assuming an array of $N_a$ disconnected antennas, the MoM impedance matrix $\mathbf{Z}$, the excitation vector $\mathbf{v}$ and the unknown coefficients $\mathbf{x}$ can be partitioned into per-antenna blocks, leading to the following system of equations:
\begin{align}
\begin{bmatrix}
\mathbf{Z}_{11}           & \dots     &   \mathbf{Z}_{1N_a}           \\
\vdots  &  \vdots    &   \vdots  \\   
\mathbf{Z}_{N_a 1}            & \dots     &    \mathbf{Z}_{N_a N_a}        \\
            \end{bmatrix}
            \begin{bmatrix}
\mathbf{x}_1     \\
\vdots  \\
\mathbf{x}_{N_a}     \\
            \end{bmatrix}
=
\begin{bmatrix}
\mathbf{v}_1     \\
\vdots  \\
\mathbf{v}_{N_a}     \\
            \end{bmatrix}
\end{align}
where a block $\mathbf{Z}_{ii}$ on the diagonal correspond the self-interaction  matrix of antenna $i$ while an off-diagonal block $\mathbf{Z}_{ij}$ contains mutual interactions between basis functions on antenna $i$ and $j$. For a perfectly electrically conducting scatterer, a MoM system of equations is derived from the electric field integral equations using Galerkin testing. The entry $mn$ of the block $\mathbf{Z}_{ij}$  corresponds to the integral reaction between testing function $\mathbf{f}_m$ on antenna $i$ and basis function $\mathbf{f}_n$ on antenna $j$. It is given by
\begin{align} 
[ \mathbf{Z}_{ij} ] _{mn}  =  \iint \limits_{S_m \times S_n}  \mathbf{f}_m(\mathbf{r}) \cdot \mathbf{G}(k,\mathbf{r},\mathbf{r}_s) \cdot \mathbf{f}_n(\mathbf{r}_s) \ \mathrm{d} S_m \mathrm{d} S_n
\label{reactionintegral}
\end{align}
where $\mathbf{G}$ is the dyadic GF describing the interaction between points $\mathbf{r}$ and $\mathbf{r}_s$ on the surfaces $S_m$ and $S_n$, respectively. 

For a large number of basis functions per antenna, filling and inverting the matrix $\mathbf{Z}$ becomes prohibitively costly even for a small number of antennas. The inversion step can be accelerated using a reduced set of Macro-Level Basis (MBF, \cite{Suter}) functions defined on the entire antenna surface and expressed as linear combinations of elementary basis functions, $\mathbf{x}_{i} \approx \mathbf{Q} \mathbf{x}_{r,i}$, with the matrix $\mathbf{Q}$ containing pre-computed MBF coefficients. This yields a reduced system of equations with the size of the reduced matrix $\mathbf{Z}_{r, ij} = \mathbf{Q}^H \mathbf{Z}_{ij} \mathbf{Q}$ being much smaller than that of the original matrix $\mathbf{Z}_{ij}$. A challenge remains in constructing the reduced matrices $\mathbf{Z}_{r, ij}$  without fully evaluating all elementary blocks $\mathbf{Z}_{ij}$ beforehand. An initial 3D multipole method was proposed in \cite{Craeyemultipole}, but it is not suitable for short distances. Another method, HARP \cite{HARP}, maintains accuracy at short distances. Assuming the same set of MBFs for each antenna, it models MBF interactions using a harmonic-polynomial function, with coefficients derived from a reduced set of elementary blocks $\mathbf{Z}_{ij}$  tabulated on a radial-azimuthal non-uniform grid \cite{Gonzalez}. As demonstrated later, pre-computing these elementary blocks can become costly for antennas with many elementary basis functions. Also, it does not allow the presence of rotated antennas in the array. In the following section, we will instead build on the first approach \cite{Craeyemultipole} and replace the HF multipole method with the SDM-based approach developed in this paper.

\subsection{Rapid evaluation of mutual interactions using the broadband SDM decomposition}
The derivation of the multipole-based MoM interaction is achieved by substituting the GF decomposition in \eqref{reactionintegral} and interchanging spatial and spectral integrations, and then re-expressing spatial derivatives as products in the spectral domain \cite{Jandhyala} \cite{Gonzalez2}. Similar to Section~\ref{subsec:3DGF}, the broadband formulation requires splitting the interactions into two terms,
\begin{align}
[ \mathbf{Z}_{ij} ] _{mn} = [ \mathbf{Z}_{ij} ] _{mn,+} + [ \mathbf{Z}_{ij} ] _{mn,-}
\label{splitZ}
\end{align}
with
\begin{align} \nonumber
 [ \mathbf{Z}_{ij} ] _{mn,\pm} & =  \frac{1}{k \eta } \int\limits_{-\infty }^{\infty} \int\limits_{0 }^{2\pi}  \  \mathbf{p}_{m}(-k_z,\alpha+\pi) \cdot  \mathbf{p}_{n}(k_z,\alpha)  \\ & T_o(k_\rho P_{ij},\pm(\alpha_{r}-\phi_{ij})+j \chi)\   \mathrm{d}\alpha_r \ k_z' \mathrm{d}k_{zr}
\label{spectralreaction}
\end{align}
where the complex pattern of basis function $m$ is given by $\mathbf{p}_{m} = \eta/(2j\lambda) (\tilde{\mathbf{f}}_{m} - (\tilde{\mathbf{f}}_{m}\cdot \mathbf{k} ) \ \mathbf{k}/k^2)$ with the Fourier spectrum $\tilde{\mathbf{f}}_{m}$ defined in \eqref{BFspectrum}. Expanding the pattern product into TE-TM components leads to
\begin{align} 
\mathbf{p}_{m}\cdot \mathbf{p}_{n} =  p_{m,TE}\cdot p_{n,TE} +  p_{m,TM}\cdot p_{n,TM}
\label{patternproduct}
\end{align}
For identical antennas, this product is independent of baseline and therefore can be pre-computed for all pairs of MBFs and stored prior to matrix filling. 

If each antenna $i$ is now rotated by an angle $\beta_i$, the pattern product \eqref{patternproduct} needs to be recalculated for each antenna pair, resulting in a slight increase in computational cost, as demonstrated later in Section~\ref{subsec:exrot}. To facilitate this rotation, we expand each patterns using azimuthal harmonics:
\begin{align} 
p_{m,q}(k_z,\alpha) =  \sum_{l=-M/2}^{M/2} c_{ml,q}(k_z) \ e^{jl\alpha_z}
\label{harmonicpat}
\end{align}
Here, $ c_{ml,q}$  are the harmonic coefficients for $TE$ and $TM$ modes, and the expansion is truncated to $\pm M$. Details on obtaining these coefficients directly from the basis function $\mathbf{f}_m$ are provided in Appendix~\ref{app:complexpat}. When antennas are rotated, \eqref{harmonicpat} is simply updated by replacing $\alpha$ with $\alpha-\beta_i$.

\begin{figure}[]
\subfloat[]{\includegraphics[trim=1.5cm 2cm 0.0cm 2cm,clip,width=10cm,height=5.5cm]{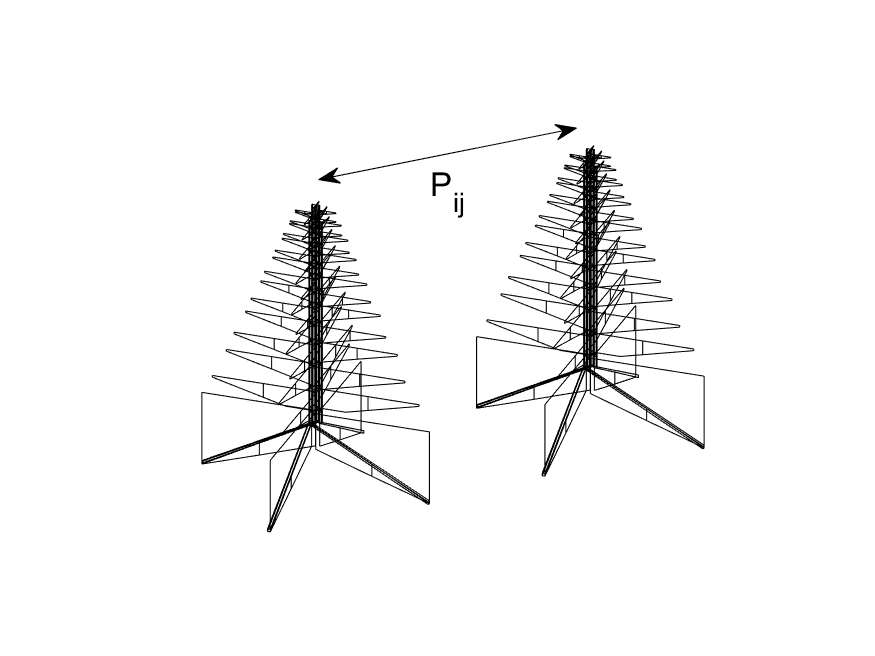}} \\
\subfloat[]{\includegraphics[trim=0.0cm 0.0cm 0.0cm 0.5cm,clip,width=9cm,height=6.5cm]{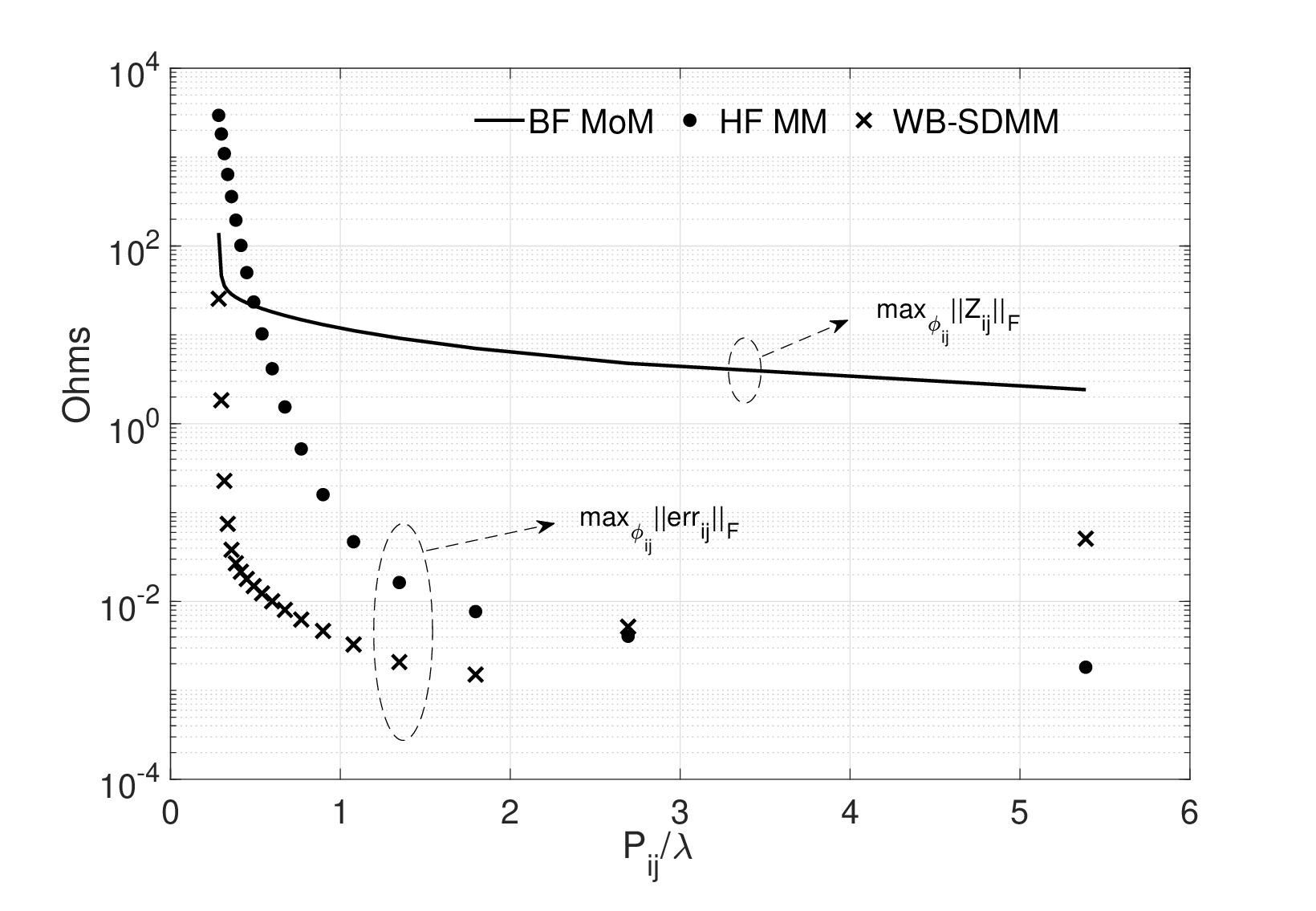}}\\
\subfloat[]{\includegraphics[trim=0.0cm 0.0cm 0.0cm 0.50cm,clip,width=9cm,height=6.5cm]{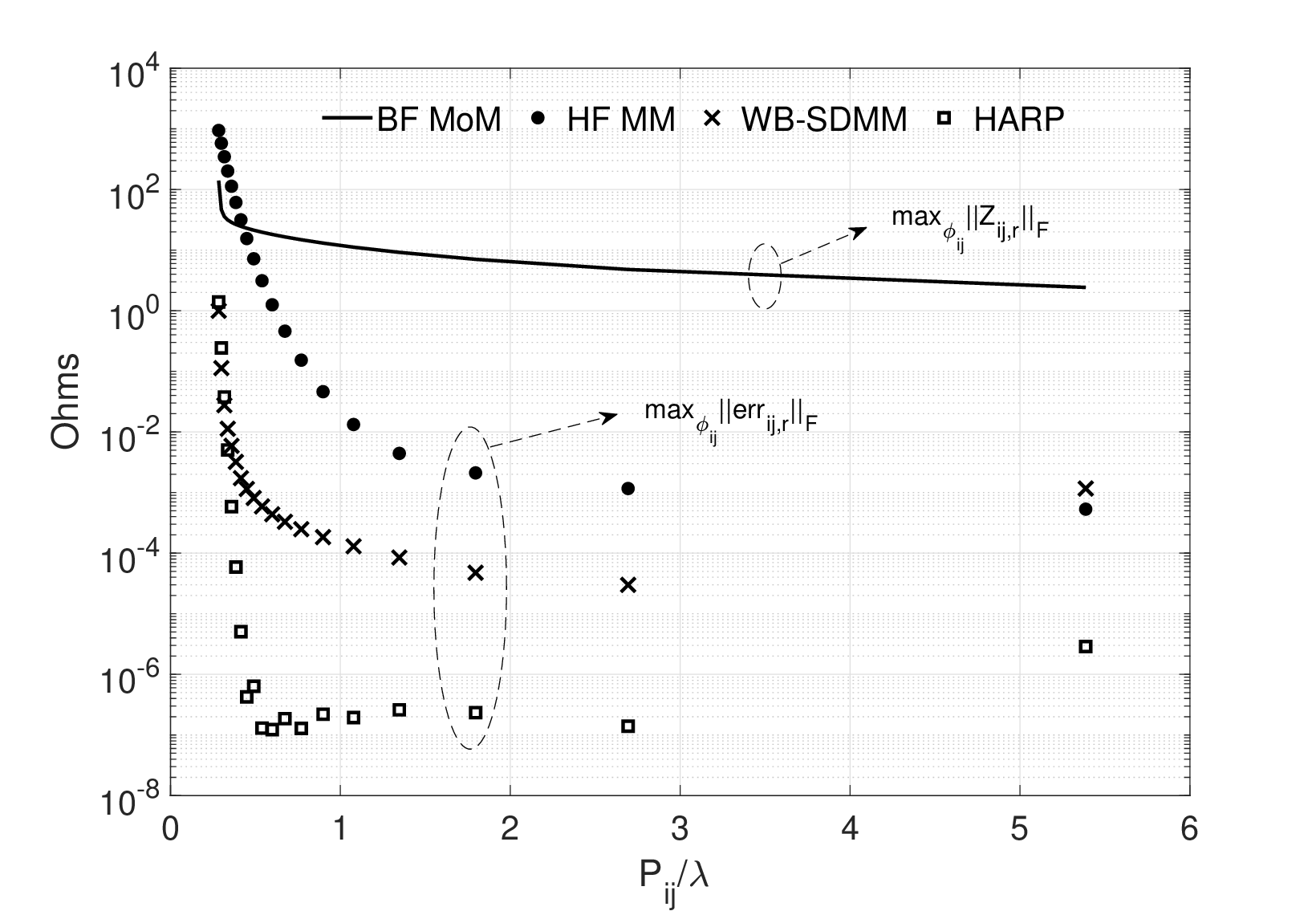}}
\caption{Numerical evaluation of MoM mutual interactions (solid line) between two SKALA4 antennas (a) using the brute-force (BF) MoM and absolute errors (markers) using the HF-MM, the proposed WB-SDMM, and HARP methods. The errors correspond to the maximum of the Frobenius norm of the absolute error on mutual interactions between elementary basis functions (b) and all $50$ MBFs (c) for all relative angles and increasing baseline lengths.}
\label{fig:MultipoleInteraction}
\end{figure}

We now validate the formulation using a single-baseline experiment involving two wideband log-periodic antennas, each standing $2$ meters tall and $1.6$ meters in diameter. This antenna, known as SKALA4 \cite{SKALA4} and shown in Fig.~\ref{fig:MultipoleInteraction}.a, is representative of those used in the Square Kilometer Array telescope \cite{SKA}. The simulations are carried out at a frequency of $50$ MHz, which is at the lower end of the operating band. Each antenna is meshed with approximately $3600$ basis functions. The minimum relative distance between the centers of the antennas is $1.7$ meters ($0.3$ wavelengths). When the dipoles of two antennas are aligned, the closest distance between metallizations is $0.1$ meters, equivalent to $0.015$ wavelengths. The baseline angle is varied from $0$ to $2\pi$. The experimental setup is illustrated in Fig.~\ref{fig:MultipoleInteraction}.a. 

We compute the absolute errors between elementary interactions using the brute-force (BF) MoM method, the wideband SDM method (WB-SDMM) with $N_z = 30$ and $M=30$, and the classical high-frequency 3D multipole method (HF-MM) with $M=10$. These errors are measured as the Frobenius norm of the difference with the BF MoM blocks. Maximum error values across all baseline angles are plotted for a given baseline length in Fig.~\ref{fig:MultipoleInteraction}.b. It is clear that the HF-MM method quickly deteriorates for distances smaller than a wavelength, whereas the WB-SDMM remains accurate, achieving 1-digit accuracy at the smallest distance. In Fig.~\ref{fig:MultipoleInteraction}.c, we compare the same errors for MBF interactions, this time including HARP with $30$ radial points and $60$ azimuthal points to parameterize the interaction model. The number of MBFs is $50$. While HARP shows slightly lower errors at intermediate and larger distances, the WB-SDMM does slightly better at small distances.
The worst-case error for MBF interactions is $2$ digits of accuracy, which is one order less than for elementary interactions.
The computation times per MBF interaction for both HARP and WB-SDMM were on the order of one-tenth of a millisecond on a laptop, demonstrating similar accuracy within comparable runtime.

\section{Numerical examples}
\label{sec:numexamples}
We now examine three different arrays and analyze computation costs and memory requirements. The simulations run on a laptop equipped with an Intel Core i7-13800H processor ($2.50$ GHz, $14$ cores) and $32.0$ GB of RAM. In these examples, we consider the dual-polarized SKALA4 antenna \cite{SKALA4}, shown in Fig.~\ref{fig:MultipoleInteraction}.a, each loaded with a $100\Omega$ resistor and operating between $50$ and $350$ MHz. The antenna is modeled using wire segments and around $3600$ wire basis functions, with the Method-of-Moments employing the thin-wire Pocklington approximation. The mutual interactions between elementary basis functions are computed in C++, though this computation is not parallelized. This in-house code has been validated against multiple commercial solvers in \cite{HARP} and in \cite{Gueuning}.
The antenna self-interaction elementary block $\mathbf{Z}_{ii}$ varies smoothly with frequency. Therefore, these can be computed and stored beforehand on a very coarse grid and subsequently rapidly interpolated at a finer frequency resolution in less than a second. For all simulations, the number of MBFs is fixed to $1$ primary and $50$ secondary MBFs and we aim to compute all the embedded element patterns (EEPs) for each antenna and for both feeds.  The EM simulator developed in this paper is named Fast Array Simulation Tool (FAST).

\subsection{Array of $256$ randomly-placed log-periodic antennas}
\begin{figure}[]
\subfloat[]{\includegraphics[trim=0.0cm 0.0cm 0.0cm 0.0cm,clip,width=4.5cm,height=3.8cm]{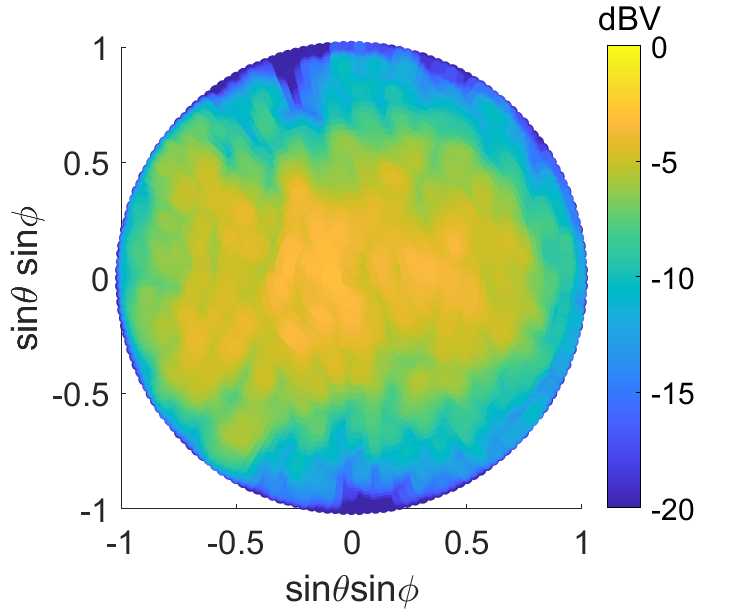}}\\
\subfloat[]{\includegraphics[trim=0.0cm 0.0cm 0.0cm 0.0cm,clip,width=4.5cm,height=3.8cm]{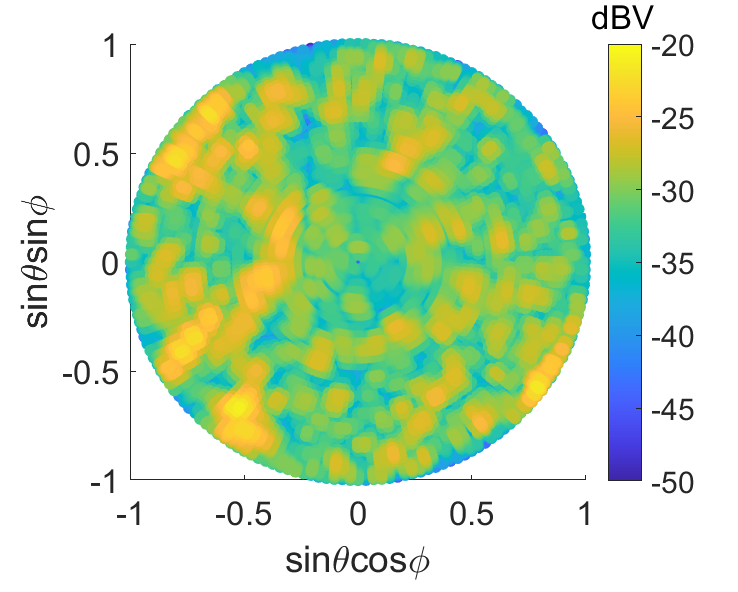}}
\subfloat[]{\includegraphics[trim=0.8cm 0.0cm 0.0cm 0.0cm,clip,width=4.5cm,height=3.8cm]{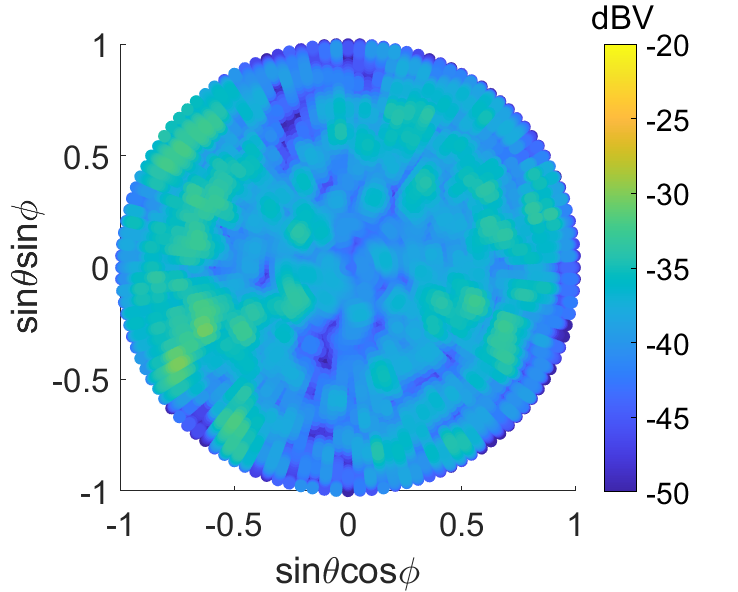}}
\caption{Example A: (a) EEP (in dBVolts) for the element squared in the right-side pseudo-random array shown in Fig.~\ref{fig:2stations} obtained with FAST. Absolute errors on the far field w.r.t. FEKO's MLFMM (b) and HARP (c) solutions.}
\label{fig:randomval}
\end{figure}

In this example, we consider an irregular array of $256$ SKALA4 antennas positioned in a pseudo-random manner (see the right-side array layout in Fig.~\ref{fig:2stations}.a). The minimum and maximum baseline lengths in this layout are $1.75$ m and $38$ m, respectively. First, we benchmark FAST and validate it against FEKO's MLFMM solver \cite{FEKO}. Note that FEKO runs on a large workstation with $128$ cores and $2$ TB of RAM and uses the MLFMM solver with a sparse LU preconditioner. The iterative solver converges in between $10$ and $20$ iterations for each EEP. We also compute the EEPs with HARP using radial and azimuthal orders $N=10$ and $M=10$. FAST is parameterized with $N_z = 14$ and $M=10$. Figure~\ref{fig:randomval}.a shows the EEP of an element near the edge of the array. The error of FAST w.r.t. FEKO (left) and HARP (right) are shown in Fig.~\ref{fig:randomval}.b and Fig.~\ref{fig:randomval}.c, respectively. The agreement with FEKO is around $-25$ dBV, while with HARP it is around $-35$ dBV, which corresponds to almost $2$ digits of accuracy in the electric far field.

\begin{table*}[]
\centering
\setlength{\tabcolsep}{3pt} 
\renewcommand{\arraystretch}{1.4} 
\centering
\caption{Summary of computation costs for the numerical examples.}
\begin{tabular}{ |p{4.8cm}|M{2.3cm}|M{2.3cm}|M{2.3cm}|M{2.3cm}|M{2.3cm}| }
\hline
Solver & MLFMM  (FEKO)  & \multicolumn{1}{c|}{HARP} & \multicolumn{3}{c|}{FAST}  \\ \hline
Machine &  AMD EPYC 7662, 128 cores, 1496 MHz, 2TB RAM & \multicolumn{4}{c|}{  Intel Core i7-13800H, 2.50 GHz, 14 cores, 32.0 GB RAM}  \\ \hline
Array configuration & \multicolumn{3}{c|}{256 random}  & 256 regular, rotated & 2 x 256 random \\ \hline
Pre-comput. time (mins) & 21 & 1313 & 10.1 & 10.1   & 10.5\\ 
Matrix filling time (mins) & - & 0.2 & 0.23 & 0.5 & 1.0 \\ 
Solve time (mins) & 230 & 0.51  & 0.51 & 0.51 & 5.9 \\ 
Total time (mins) & 251 & 1314 & 10.8 & 11.1 & 17.4 \\ 
Re-comput. time per new layout (mins) & 251 & 0.71 & 0.74 & 1.01 & 6.9 \\ 
Peak Memory (GBs) & 359 & 1.3 & 1.3 & 1.3 & 5.3 \\
\hline
\end{tabular}
\label{table:computationtimes}
\end{table*}

The computation times and peak memory usage are summarized in Table~\ref{table:computationtimes}. The preparation time for FAST is about $100$ times faster than for HARP. Moreover, FAST running on a laptop outperforms FEKO's MLFMM on a $128$-cores workstation by a factor of $25$ in total run time. Both FAST and HARP use only $3.7$ GB of RAM, which is significantly less than the $350$ GB required by the MLFMM. Modifying the array layout with HARP and FAST involves only re-filling and re-solving the reduced MoM matrix, taking approximately one additional minute. In contrast, the MLFMM requires a complete restart, consuming an additional $4$ hours.
Regarding pre-computations, HARP spends most of its time computing elementary blocks for MBF generation and harmonic-polynomial model parametrization, requiring $50$ and $200$ MoM blocks, respectively, each taking under $6$ minutes on a laptop. In comparison, FAST takes around $10$ seconds to generate $1$ MBF and its complex pattern. Reduced matrix filling takes about $12$ seconds for HARP and $14$ seconds for FAST. Solving the $12800\times12800$ reduced MoM matrix takes $30$ seconds for both methods.
In conclusion, FAST's pre-computation steps are around $2$ orders of magnitude faster than HARP’s, while both methods have comparable times for re-computation with a new layout. This speed-up may however vary depending on the number of basis functions per antennas.

\subsection{Regular array of $256$ randomly-rotated antennas}
\label{subsec:exrot}

\begin{figure}[]
\subfloat[]{ \includegraphics[trim=0.0cm 0cm 0.0cm 0cm,clip,width=9.5cm,height=4.5cm]{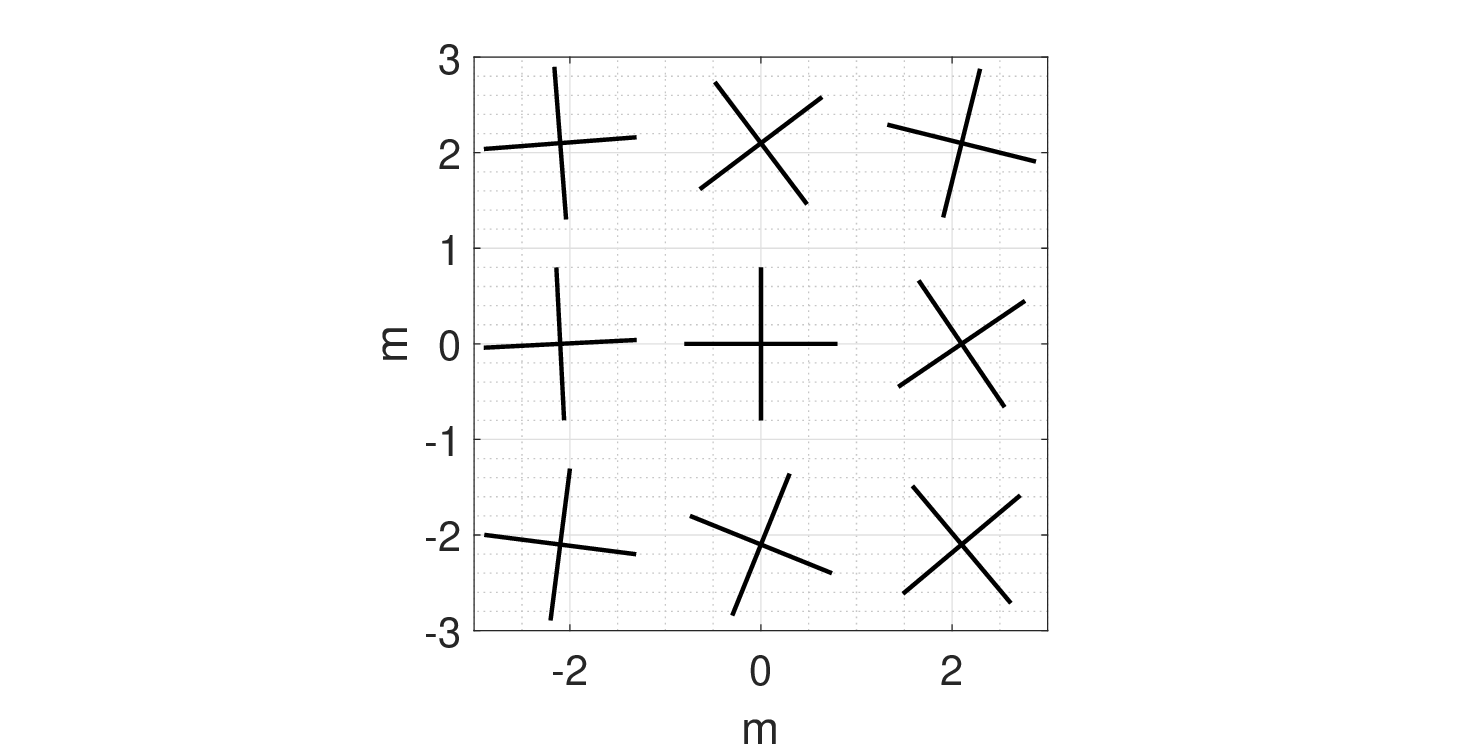}}\\
\subfloat[]{\includegraphics[trim=0.0cm 0cm 0.0cm 0cm,clip,width=9.5cm,height=5.2cm]{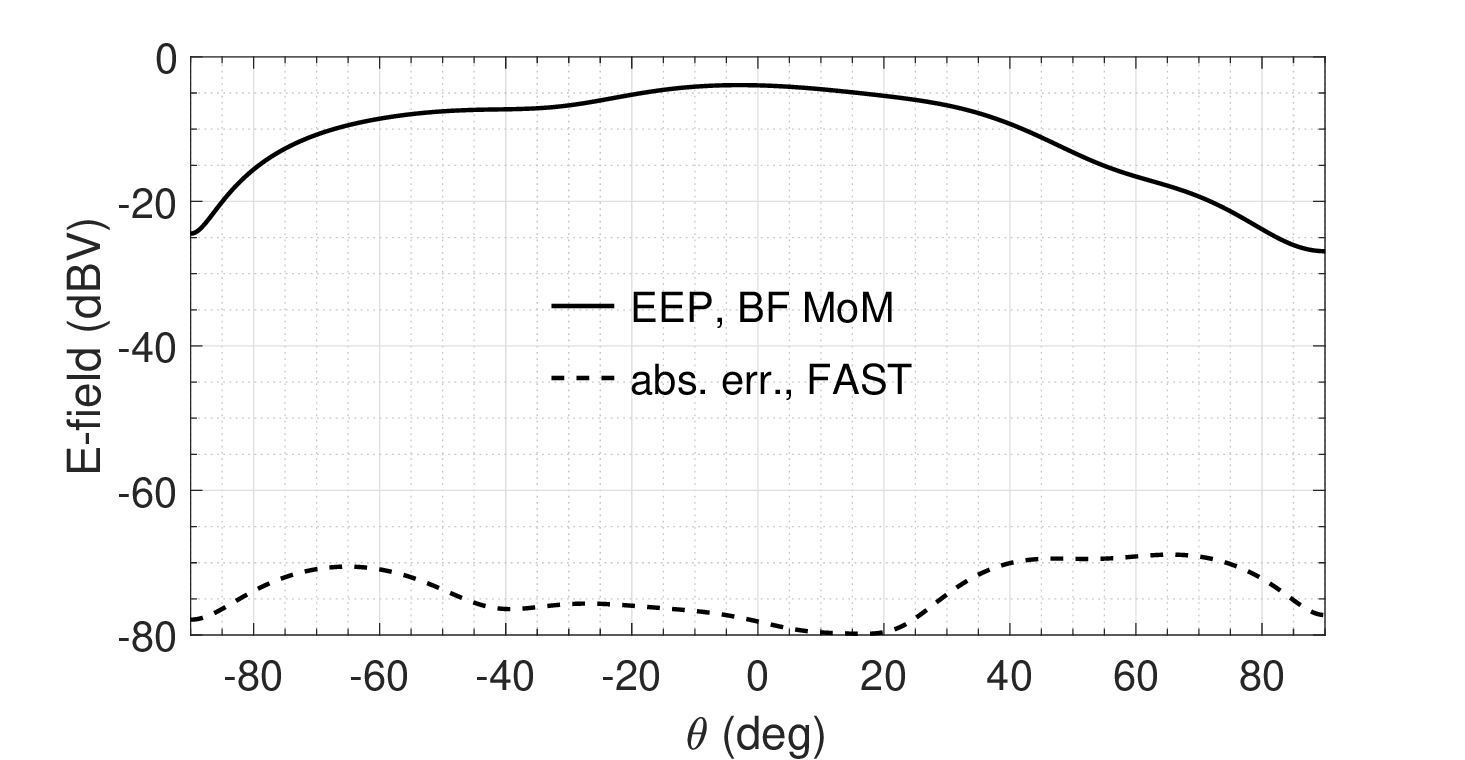}}
\caption{Example B: Validation of FAST against BF MoM method for rotated antennas in a $3\times3$ regular array with $2.14$m inter-element distance (a), simulated at $125$ MHz. The cross-section in the EEP X-plane (b) demonstrates $3-4$ digits agreement in the electric far field.}
\label{fig:validationrot}
\end{figure}

For rotated antennas, we use the same set of MBFs as in the unrotated configuration. To validate this assumption, we conducted a small example with a $3\times3$ regular array with a $2.1$ m inter-element distance, simulated at $125$ MHz. As shown in Fig.~\ref{fig:validationrot}, FAST achieves agreement with the BF MoM method down to approximately $-75$ dBV in the electric far field, indicating almost $4$ digits agreement. 

\begin{figure}[]
\subfloat[]{\includegraphics[trim=0.0cm 0cm 0.0cm 0cm,clip,width=9.5cm,height=5.2cm]{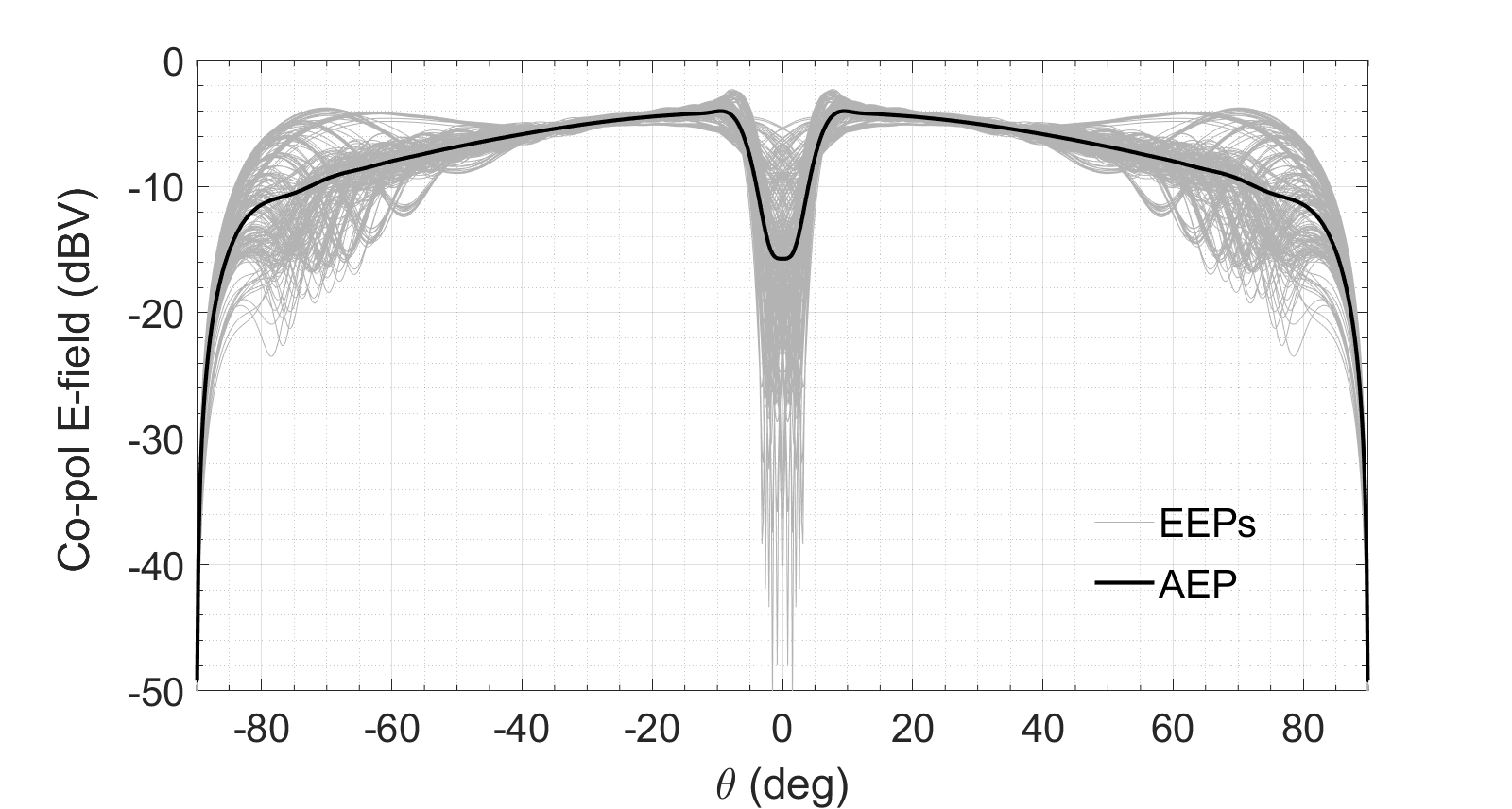}}\\
\subfloat[]{\includegraphics[trim=0.0cm 0cm 0.0cm 0cm,clip,width=9.5cm,height=5.2cm]{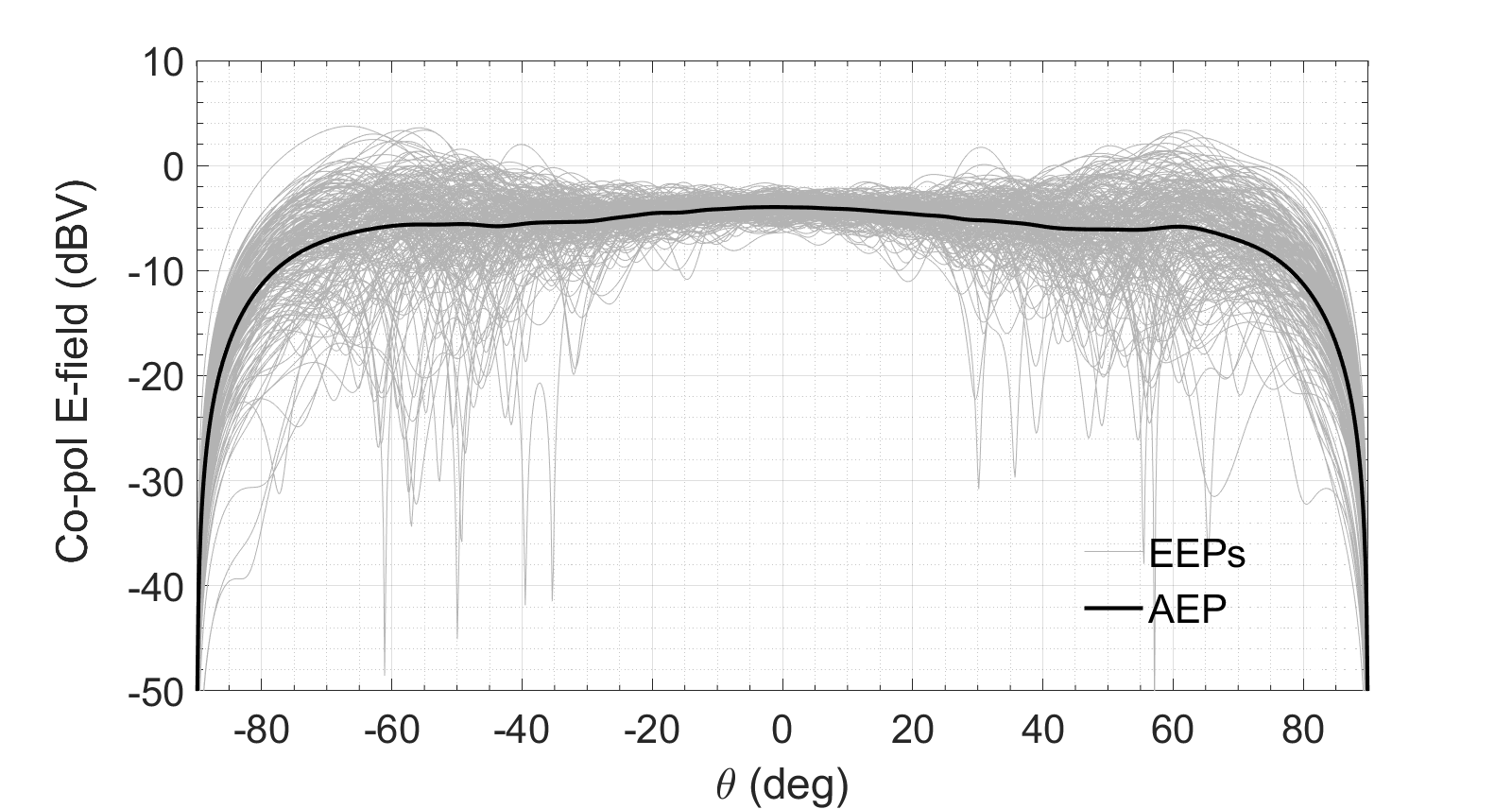}}
\caption{Example B: Mitigation of a zenith blindness effect in the Y-plane of X-feed EEPs in a regular $16\times16$ array with $2.14$m separation at 125 MHz, without (a) and with (b) random antenna rotations.}
\label{fig:dip125}
\end{figure}

We will now extend the study to a larger array with $16\times16$ elements. Cuts of the $256$ EEPs for feed X in the Y-plane are depicted in Fig.~\ref{fig:dip125}.a. One immediately notes a significant drop of power around zenith. This effect has been evidenced in simulations in \cite{Davidson}\cite{Anstey} and results from destructive interference from coupled elements in the regular layout. One way to mitigate this effect is to randomly perturb the antenna positions \cite{Anstey}. Another mitigation approach involves individually rotating each antenna by a random angle while maintaining the layout unchanged. With this strategy, the received voltages from the antennas must be back-rotated in post-processing to align the antenna polarisations,
\begin{align}
            \begin{bmatrix}
v_{X,i}     \\
v_{Y,i}    \\
            \end{bmatrix}
=
\begin{bmatrix}
\cos\beta_i              &   -\sin \beta_i          \\
\sin\beta_i      &    \cos\beta_i        \\
            \end{bmatrix}
\begin{bmatrix}
v_{X_i,i}     \\
v_{Y_i,i}     \\
\end{bmatrix}
\end{align}
where indices $X_i$ and $Y_i$ refer to antenna $i$'s rotated feed orientations while $X$ and $Y$ refer to the fixed reference axes. As shown in Fig.~\ref{fig:dip125}.b, the dip in the Y-plane of the back-rotated EEPs has disappeared, albeit with an increase in variance between elements. Further investigations are necessary to ensure that this correction does not introduce spurious features elsewhere in the frequency band or across the field of view. This design issue is beyond the scope of this analysis paper.

The computation times for this example are summarized in Table~\ref{table:computationtimes}. It is noted that compared to the un-rotated case, the matrix filling step takes $2$ times longer due to the need to re-evaluate the pattern product \eqref{patternproduct} for each baseline.

\subsection{Two adjacent stations of the SKA-Low telescope} 
\begin{figure}[]
\subfloat[]{\includegraphics[trim=0.0cm 0.0cm 0.0cm 0.5cm,clip,width=9cm,height=5cm]{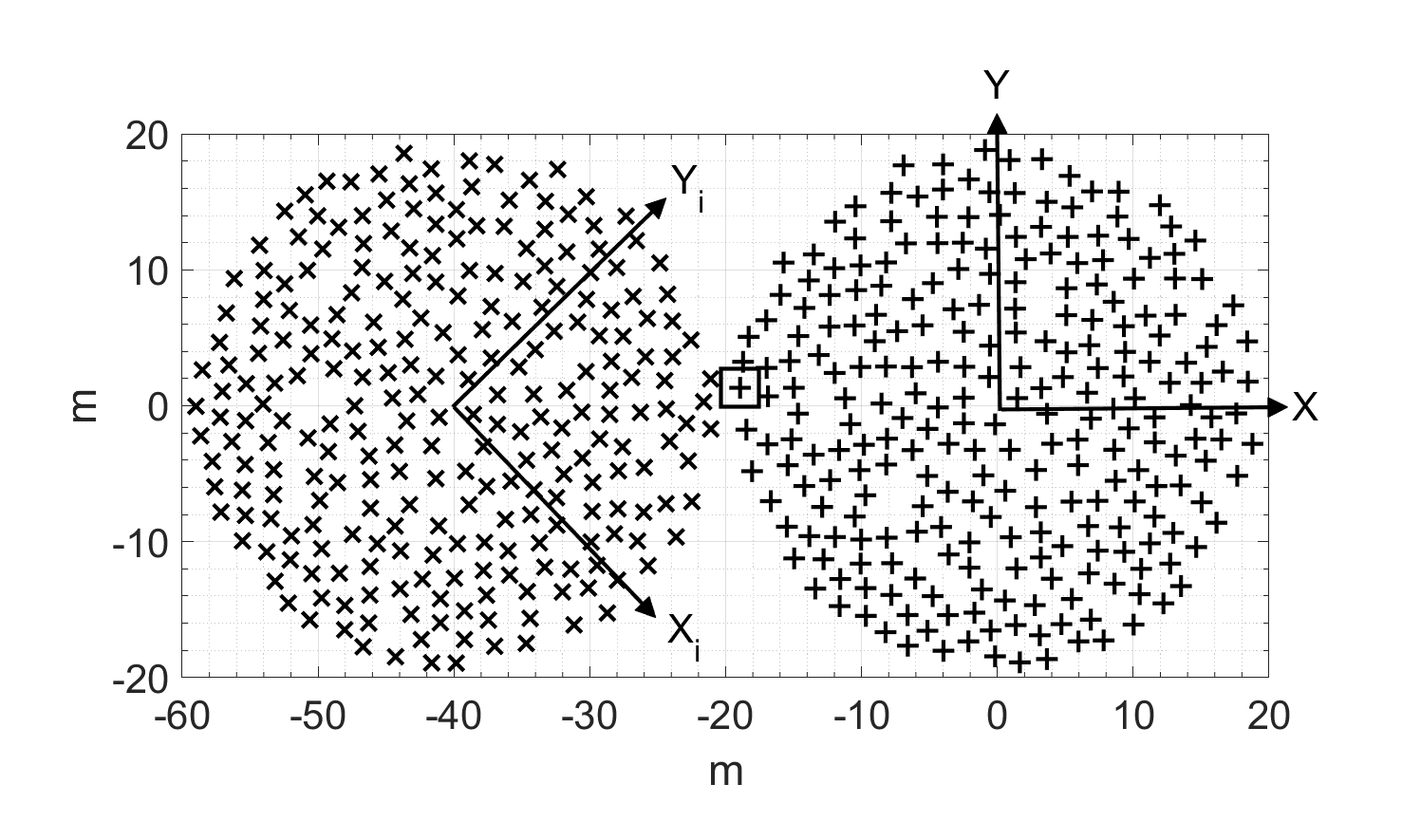}}\\
\subfloat[]{\includegraphics[trim=0.0cm 0.0cm 0.0cm 0.5cm,clip,width=9cm,height=5cm]{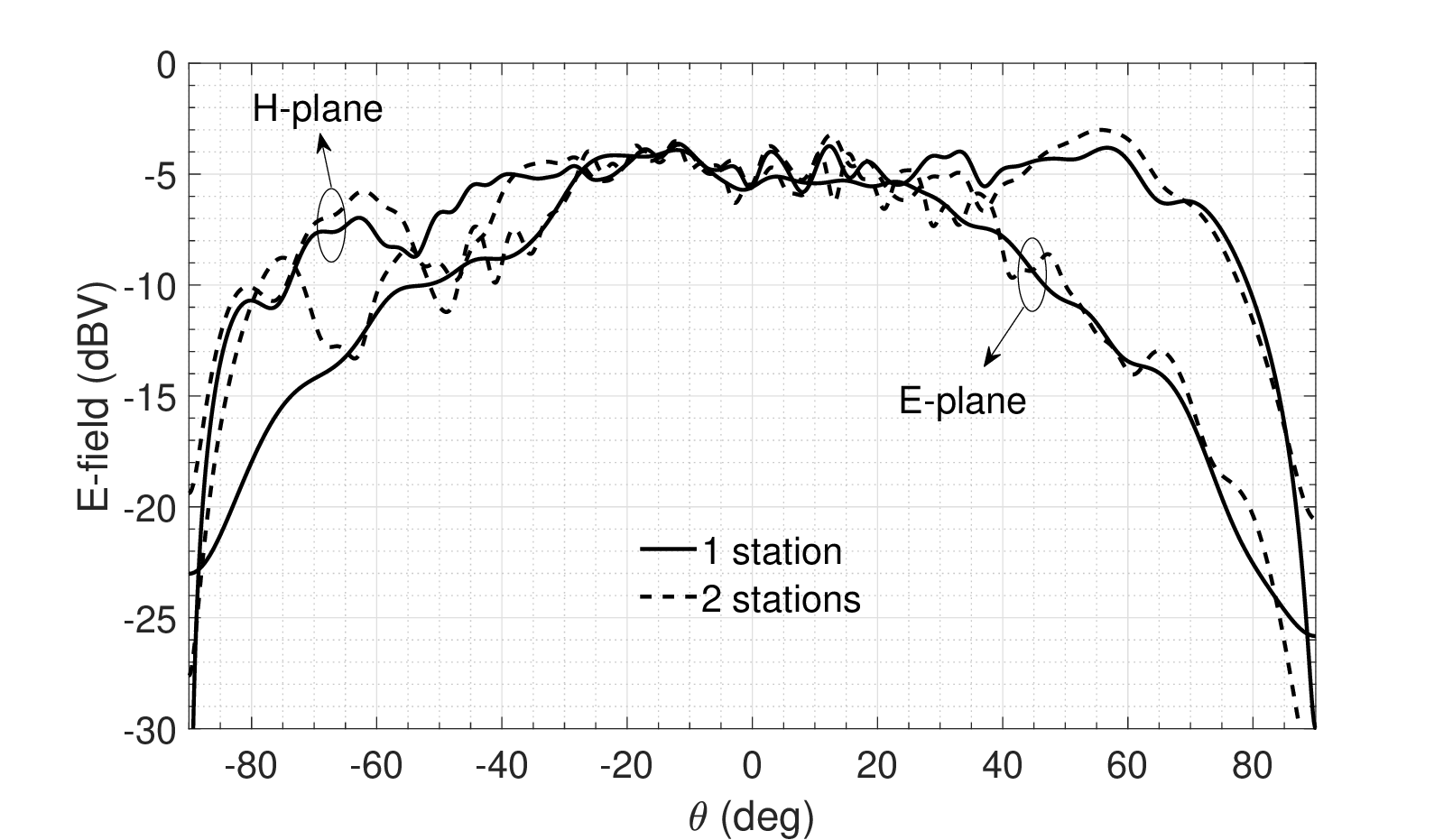}}
\caption{Example C: Two close SKA-Low stations (a) with random layout and with the left-side station rotated by $45$ degrees. The EEP (b) for an edge element (squared) at 125 MHz is shown in the E- and H-plane for feed X with and without the second nearby station.}
\label{fig:2stations}
\end{figure}

The final example involves simulating two closely located phased arrays, called stations, of the low-frequency SKA telescope (SKA-low,\cite{SKA}). In the core of the telescope, stations are densely packed, with minimum distances between them comparable to those within a single array. A devised design approach for SKA-low design is to rotate each of the $512$ stations by a given angle varying from $0$ to $2\pi$. Considering the same random array layout as in Example A, the scenario is illustrated in Fig.~\ref{fig:2stations}. The simulation frequency is set at $125$ MHz. The EEP for an edge element (squared) is plotted and compared to that of example A, which included only the right-hand side station. Significant deviations in the EEP near the horizon and more pronounced ripples are observed. Regarding computation times, this case is a hybrid between examples A and B, as antennas within a station are not rotated relative to each other, but there is a single relative rotation angle between the stations. The $2$ pattern products, with and without this relative rotation, can be pre-computed beforehand. Consequently, the filling time is not increased compared to an equivalent un-rotated scenario.

\section{Conclusion}
\label{sec:Conclusion}
This paper introduces a numerical method for analyzing mutual coupling effects in large, dense, and irregular arrays of identical antennas, each with potentially thousands of basis functions. The method leverages the Method of Moments (MoM) with a Macro Basis Function (MBF) approach for rapid direct inversion of the MoM impedance matrix, and a Steepest-Descent Multipole Method (SDMM) to accelerate the matrix filling step. Key advancements include:
\begin{enumerate}
\item  Broadband SDMM: We have introduced a stable and efficient broadband expansion for quasi-planar problems, covering low to high frequencies. This approach combines steepest-descent path integration on the vertical axis with a 2D multipole decomposition. It is more efficient than classical 3D multipole methods at high frequencies, requiring only plane waves propagating near the horizontal plane. Stabilization at low frequencies is achieved with an imaginary shift in the angular integral, and efficiency is further improved by reducing the number of required evanescent waves through a non-uniform sampling scheme. To the author's knowledge, this is the first stable and efficient multipole-based expansion presented for quasi-planar structures.
\item Low pre-computation costs: the SDMM requires only the pre-computation of each MBF's complex patterns, resulting in lower antenna-dependent pre-processing costs compared to existing techniques. This is particularly crucial in case of wideband antennas meshed with many elementary basis functions.
\item Support for rotated antennas: the method accommodates arrays with arbitrarily rotated antennas without incurring significant additional computational cost, offering flexibility for practical applications where antenna orientation may vary. 
\end{enumerate}
Examples showed that EM simulations with irregular arrays comprising $256$ wideband log-periodic antennas can be completed in just $10$ minutes on a current laptop, with an additional minute to re-compute the EEPs for a new layout. This corresponds to speed-up factors of $25$ and $250$, respectively, compared to FEKO's MLFMM running on a much larger workstation. These lower computation times enable the optimization of both antenna geometry, orientation and array layout, as well as sweeping at fine frequency resolution. The broadband SDM expansion could also find applications in the scattering analysis of rough surfaces or planar arrays of dissimilar and/or connected antennas through its implementation into an iterative solver.

\appendices
\section{Harmonic expansion of complex patterns}
\label{app:complexpat}

Let us start with the Fourier spectrum of basis function $m$,
\begin{align} 
\tilde{\mathbf{f}}_{m}(\mathbf{k})   = \iint \limits_{S_m}  \mathbf{f}_m(\mathbf{r})  \ e^{-j\mathbf{k}\cdot (\mathbf{r}-\mathbf{r_j})} \ \mathrm{d}S_m
\label{BFspectrum}
\end{align}
Then, substituting the Jacobi-Anger expansion of the plane-wave exponential \cite{Abramowitz} leads to
\begin{align} 
\tilde{\mathbf{f}}_{m}(k_z, \alpha)   = \sum_{n=-\infty}^{\infty} \mathbf{a}_{mn}(k_z,\alpha) \ e^{jn\alpha}
\end{align}
with the vector coefficients given by 
\begin{align} 
\mathbf{a}_{mn}(k_z,\alpha) = \iint \limits_{S_m}  \mathbf{f}_m(\mathbf{r}) \ J_n(k_\rho \rho) \ e^{-jn\phi}\ e^{-jk_z z} \ \mathrm{d}S_m
\end{align}
where $J_n$ is the first-kind Bessel's function of order $n$.
Given that the complex pattern reads $\mathbf{p}_{m} = \eta/(2j\lambda) (\tilde{\mathbf{f}}_{m} - (\tilde{\mathbf{f}}_{m}\cdot \mathbf{k} ) \ \mathbf{k}/k^2)$ and the TE and TM polarisation vectors $\hat{\mathbf{e}}_{TE}  = \hat{\mathbf{z}} \times \mathbf{k}/k_\rho $ and $\hat{\mathbf{e}}_{TM}  = \hat{\mathbf{e}}_{TE} \times \mathbf{k}/k$, we can derive the harmonic coefficients for TE-TM components of the complex pattern,
\begin{align} 
p_{m,q}(k_z, \alpha)   = \sum_{n=-\infty}^{\infty} c_{mn,q}(k_z,\alpha) \ e^{jn\alpha}
\end{align}
with $q = \{TE,TM\}$ and
\begin{align} \nonumber
c_{mn,TE} & = k_z/(2k)  (a_{n+1,x} + a_{n-1,x} + ja_{n+1,y} \\  \nonumber &  -ja_{n-1,y}) - a_{n,z}k_{\rho}/k \\ \nonumber
c_{mn,TM} & = 1/2(-ja_{n+1,x} + ja_{n-1,x} + a_{n+1,y} \\   & + a_{n-1,y})
\end{align}

\section*{Acknowledgment}
This research was supported by the Science and Technology Facilities Council (STFC). Q. Gueuning was supported by grant number
ST/W00206X/1 and E. de Lera Acedo acknolewdges STFC for their support via a Ernest Rutherford Fellowship (ERF).

\begin{IEEEbiography}[{\includegraphics[width=1in,height=1.25in,clip,keepaspectratio]{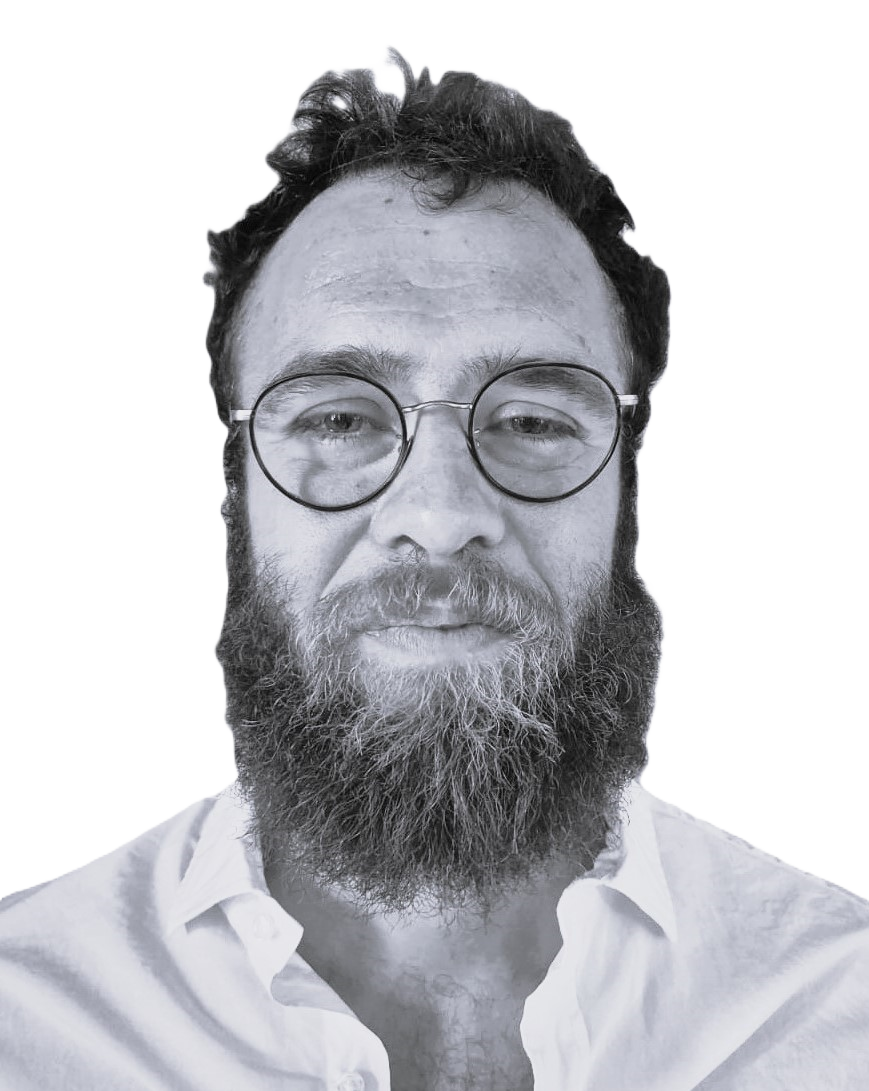}}]{Quentin Gueuning} 
received his Ph.D. from the Université catholique de Louvain (UCLouvain), Louvain-la-Neuve, Belgium, in 2019. He is currently a Senior Research Associate at the University of Cambridge and a member of the Astrophysics Group and the Kavli Institute for Cosmology at the Cavendish Laboratory, University of Cambridge, UK. He is also a member of an Agile Release Train, contributing to software development efforts for the Square Kilometer Array telescope.

His current research interests include fast computational methods, asymptotic high-frequency theories, mutual coupling analysis, array and antenna design and measurements, and the calibration and forward modeling of radio telescopes.
\end{IEEEbiography}

\begin{IEEEbiography}[{\includegraphics[width=1in,height=1.25in,clip,keepaspectratio]{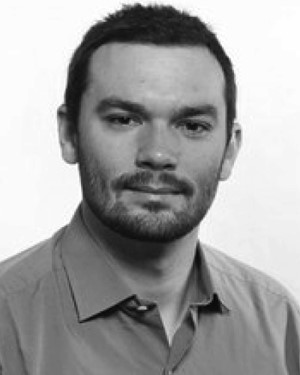}}]{Eloy de Lera Acedo} received the Ph.D. degree from
the University Carlos III of Madrid, Getafe, Spain, in 2010. He is currently the PI and a Principal Research Associate with the Cavendish Astrophysics Group, where he has developed his career on 21 cm radio cosmology and instrumentation for radio astrophysics. He is also the PI of the REACH experiment, a project aiming to detect the 21 cm global signature signal from the Cosmic Dawn and the Epoch of Reionization. He leads the Cm-wave Radio Astronomy and Novel Sensors Research Group, Cavendish Astrophysics, Cambridge, U.K., and has led the Antenna and low noise ampliﬁer (LNA) Work package (AL WP) of the Aperture Array Design Consortium, which he developed the array stations for the SKA1-LOW telescope. Within the AL WP, he has also led the design and development of SKALA, and the SKA Log periodic Antenna and its LNA. He has also led substantial pieces of work in other international projects such as the Hydrogen Epoch of Re-ionization Array (HERA) for which he has led the design of the wide band feed within the analog team group. His research interest ranges from ultra wideband array and antenna design, electromagnetic modeling, telescope calibration, and lownoise systems to the broader science case of 21 cm cosmology experiments.
\end{IEEEbiography}

\begin{IEEEbiography}[{\includegraphics[width=1in,height=1.25in,clip,keepaspectratio]{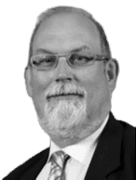}}]{ Anthony K. Brown} (Life Senior Member, IEEE) has been working in the field of antennas and since 1974.  He is currently Professor of Radiative Electromagnetics at Queen Mary, University of London, and Professor Emeritus (Communications Engineering) at the University of Manchester, UK. He was the Head of the School of Electrical and Electronic Engineering at the University of Manchester, UK until 2016 and is a former Associate Dean for Teaching and Learning at the same University. He founded EASAT Antennas Ltd (now Easat Radar Systems) in 1987 and has been CTO, CEO and  Company Chairman.  

His research in antennas and propagation has resulted in over 150 publications including five patents and, as co-author, two books.  He now concentrates on the applications of broad band antennas to radar, communications, and radio astronomy and on the characterization of large electromagnetic structures. He is a Life Senior Member of the IEEE, and a Fellow of both the Institution of Engineering Technology and the Institution of Mathematics and its Applications. 
\end{IEEEbiography}

\begin{IEEEbiography}[{\includegraphics[width=1in,height=1.25in,clip,keepaspectratio]{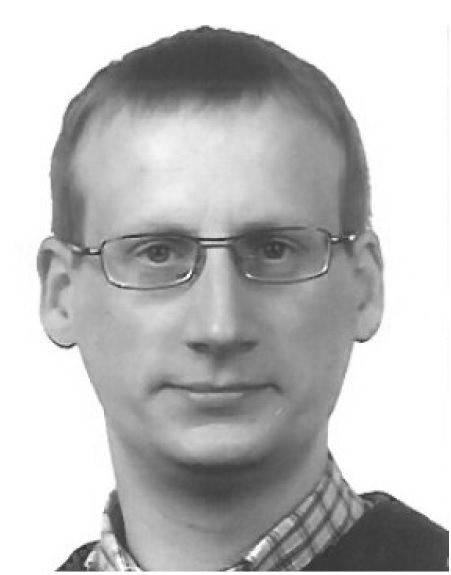}}]{Christophe Craeye} (Senior Member, IEEE) was born in Belgium in 1971. He received the Electrical Engineering and B.Phil. degrees from the Université catholique de Louvain (UCL), Louvain-la-Neuve,
Belgium, in 1994, and the Ph.D. degree in applied sciences from UCL in 1998.
From 1994 to 1999, he was a Teaching Assistant with UCL and carried out research on the
radar signature of the sea surface perturbed by rain, in collaboration with NASA and the European
Space Agency (ESA). From 1999 to 2001, he was a Post-Doctoral Researcher 
with Eindhoven University of Technology,
Eindhoven, The Netherlands. He was with the University of Massachusetts, 
Amherst, MA, USA, in 1999. He was with Netherlands Institute for Research
in Astronomy, Dwingeloo, The Netherlands, in 2001. In 2002, he started
an antenna research activity with UCL, where he is currently a Professor.
He was with the Astrophysics and Detectors Group, University of Cambridge,
Cambridge, U.K., in 2011. His research was funded by the Region Wallonne,
the European Commission, ESA, Fonds National de la Recherche Scientifique
(FNRS), and UCL. His current research interests include mutual coupling,
finite antenna arrays, wideband antennas, small antennas, metamaterials, fast
physical optics, and numerical methods for fields in periodic media, with
applications to communication and sensing systems.
Dr. Craeye received the 2005–2008 Georges Vanderlinden Prize from the
Belgian Royal Academy of Sciences in 2009. He was an Associate Editor of
the IEEE TRANSACTIONS ON ANTENNAS AND PROPAGATION from 2004 to
2010 and the IEEE ANTENNAS AND WIRELESS PROPAGATION LETTERS
from 2011 to 2017.
\end{IEEEbiography}

\begin{IEEEbiography}[{\includegraphics[width=1in,height=1.25in,clip,keepaspectratio]{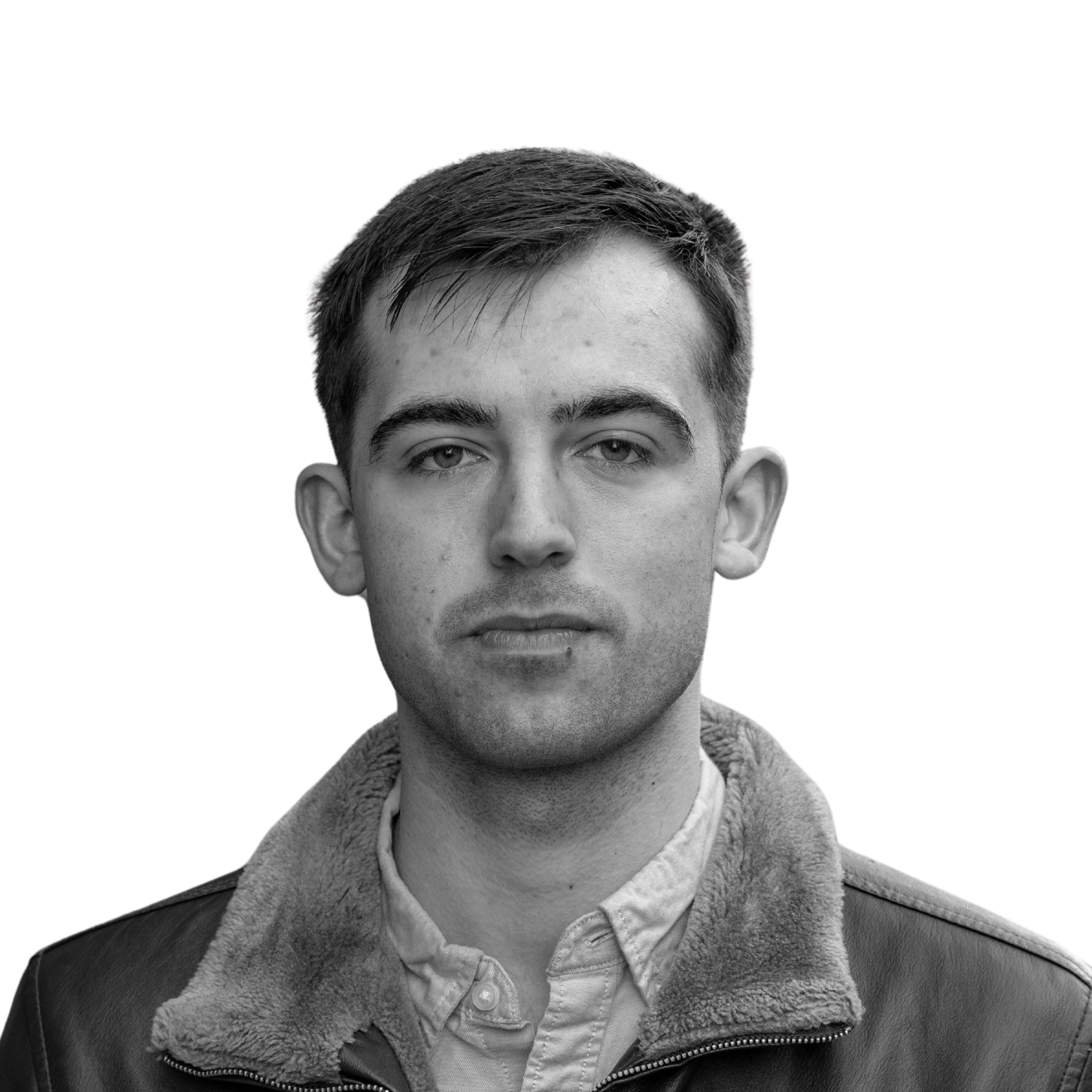}}]{Oscar} 
Oscar Sage David O’Hara was born in Oxfordshire, United Kingdom, in 1999. He received the B.A. degree in physic from Trinity College Dublin, Ireland, in 2021 and is currently pursuing a Ph.D. degree at the Cavendish Laboratory, University of Cambridge. From 2019 to 2021, he was a part time Research Assistant with The Dublin Institute for Advanced Studies covering the very low-frequency monitoring of the Earth's Ionosphere and the impacts of solar flares.

Mr. O’Hara’s current research includes the development of a high sensitivity Rydberg atomic sensor and the simulated analysis of potential systematics for the Square Kilometre Array.
\end{IEEEbiography}

\end{document}